\documentclass[twocolumn]{aastex631}

\usepackage{natbib}
\usepackage{enumerate}
\usepackage{multirow}
\usepackage{color}
\usepackage{url} 

\newcommand{\lya}{Ly$\alpha\ $}

\graphicspath{{./figures/}}

\begin{document}

\title{The Magellan M2FS spectroscopic survey of high-redshift galaxies: the brightest Lyman-break galaxies at $z \sim 6$}

\author[0000-0003-0964-7188]{Shuqi Fu}
\affiliation{Department of Astronomy, School of Physics, Peking University, Beijing 100871, China}
\affiliation{Kavli Institute for Astronomy and Astrophysics, Peking University, Beijing 100871, China}

\author[0000-0003-4176-6486]{Linhua Jiang}
\affiliation{Department of Astronomy, School of Physics, Peking University, Beijing 100871, China}
\affiliation{Kavli Institute for Astronomy and Astrophysics, Peking University, Beijing 100871, China}

\author[0000-0001-9442-1217]{Yuanhang Ning}
\affiliation{Department of Astronomy, Tsinghua University, Beijing 100084, China}

\author[0000-0002-4385-0270]{Weiyang Liu}
\affiliation{Department of Astronomy, School of Physics, Peking University, Beijing 100871, China}
\affiliation{Kavli Institute for Astronomy and Astrophysics, Peking University, Beijing 100871, China}

\author[0000-0003-0230-6436]{Zhiwei Pan}
\affiliation{Department of Astronomy, School of Physics, Peking University, Beijing 100871, China}
\affiliation{Kavli Institute for Astronomy and Astrophysics, Peking University, Beijing 100871, China}

\begin{abstract}

We present a study of a sample of 45 spectroscopically confirmed, UV luminous galaxies at $z\sim 6$. They were selected as bright Lyman-break galaxies (LBGs) using deep multi-band optical images in more than 2 deg$^2$ of the sky, and subsequently identified via their strong \lya emission. The majority of these LBGs span an absolute UV magnitude range from $-22.0$ to $-20.5$ mag with \lya equivalent width (EW) between $\sim$10 and $\sim$200 \AA, representing the most luminous galaxies at $z\sim 6$ in terms of both UV continuum emission and \lya line emission. 
We model the SEDs of 10 LBGs that have deep infrared observations from HST, JWST, and/or Spitzer, and find that they have a wide range of stellar masses and ages. They also have high star-formation rates ranging from a few tens to a few hundreds of Solar mass per year. 
Five of the LBGs have JWST or HST images and four of them show compact morphology in these images, including one that is roughly consistent with a point source, suggesting that UV luminous galaxies at this redshift are generally compact. The fraction of our photometrically selected LBGs with strong \lya emission ($\mathrm{EW}>25$~\AA) is about $0.2$, which is consistent with previous results and supports a moderate evolution of the IGM opacity at the end of cosmic reionization. Using deep X-ray images, we do not find evidence of strong AGN activity in these galaxies, but our constraint is loose and we are not able to rule out the possibility of any weak AGN activity.

\end{abstract}


\section{Introduction} \label{sec:intro}

High-redshift ($z\geq 6$) galaxies are a key probe to understand the early Universe, including the early galaxy formation and evolution, the development of large-scale structures, and the history of cosmic reionization. In the past twenty years, a number of works have been conducted to search for high-redshift galaxies, and the number of known galaxies has increased dramatically thanks to the advances of instrumentation on the Hubble Space Telescope (HST) and large ground-based telescopes (e.g., \citealt{kashikawa_2006_subaru,Vanzella_2009_VLT,hu_atlas_2010,Wilkins_2010_hstLBGz7,Stark_2011_Keck,mcleod_z_2016,bouwens_new_2021,zheng_first_2017,ono_great_2018_GOLDRUSH,ning_magellan_2020,Wold_2022_lager}). The majority of the earlier known galaxies at $z \geq 6$ are photometrically selected Lyman-break galaxies (LBGs) or LBG candidates selected by the dropout technique. The narrowband technique (or the \lya technique) is a complementary method to find high-redshift galaxies. Ground-based narrowband imaging surveys are efficient in finding \lya emitting galaxies (\lya emitters, or LAEs) at certain redshift slices such as $z\approx5.7$, 6.6, and 7.0 that correspond to wavelength windows with weak OH sky lines. 

Before the launch of James Webb Space Telescope (JWST), the majority of the photometrically selected high-redshift galaxies have not been spectroscopically observed. While narrowband selected LAE candidates typically have a high success rate of spectroscopic confirmation, only a small fraction of photometrically selected LBGs are spectroscopically confirmed (e.g., \citealt{toshikawa_discovery_2012,finkelstein_galaxy_2013,watson_dusty_2015,roberts-borsani_z_2016,song_keckmosfire_2016,larson_searching_2022}). Spectroscopic observations are important to understand LBGs, because they can not only tell us whether an object is a real  LBG, but also provide a key parameter, redshift. 
Large samples of high-redshift galaxies have been frequently used to measure galaxy properties, such as UV slopes (e.g., \citealt{dunlop_critical_2012,finkelstein_candels_2012-1,bouwens_uv-continuum_2014}), galaxy morphology (e.g., \citealt{guaita_lyman_2015,kawamata_sizes_2015,shibuya_morphologies_2015,shibuya_morphologies_2016,curtis-lake_non-parametric_2016,kobayashi_morphological_2016,liu_dark-ages_2017,Naidu_2022}), stellar populations, and star-formation rates (e.g., \citealt{egami_spitzer_2005,stark_keck_2013,gonzalez_slow_2014,faisst_coherent_2016,castellano_optical_2017,karman_muse_2017}).  These studies are mostly based on photometric samples. 

Recently, JWST is revolutionizing our understanding of high-redshift galaxies. JWST is not only breaking redshift records for the most distant objects \citep[e.g.,][]{Arrabal_Haro_2023_2z10GLX,Curtis_Lake_2023_4z10_13GLX,Fujimoto_2023_11z9_13GLXCEERS,Hsiao_2023_z10GLX_merger,Roberts_Borsani_2023_z10GLX}, but also rapidly expanding the sample sizes of high-redshift galaxies, photometrically and spectroscopically (e.g., \citealt{Atek_2023_jwst_dropout,Champagne_2023_jwst_dropout,Leethochawalit_2023_jwst_dropout,Yan_2023_jwst_dropout_2}).
In particular, JWST NIRCam and NIRISS observations offer slitless spectroscopy that makes it feasible to obtain spectra for large galaxy samples (e.g., \citealt{Matthee_2023_EIGER,Oesch_2023_grism_survey,Wang_2023_ASPIRE}). In addition, JWST NIRSpec grating observations provide medium-resolution spectra that allow detailed studies of line emission and absorption (e.g., \citealt{Bunker_2023_jades,Mascia_2023_glass,Prieto-Lyon_2023_NIRSpec,Tang_2023_ceersspec,Tacchella_2023_NIRSpec}). Despite the great power of JWST, its field-of-view is small compared to ground-based telescopes. Therefore, large-area, ground-based observations are highly complementary to JWST observations.

High-redshift galaxies are also thought to be the major source of ionizing photons at the epoch of cosmic reionization. Recent studies have suggested that star-forming galaxies, rather than quasars or AGN, provide most of the ionizing photons needed for reionization (e.g., \citealt{finkelstein_candels_2012,robertson_cosmic_2015,onoue_minor_2017,jiang_definitive_2022}). 
Spectroscopic samples of high-redshift LBGs can be used to constrain the properties of the IGM during this epoch. For example, \lya photons are easily absorbed by neutral hydrogen, making the \lya line a sensitive probe of the ionization state of the IGM. Therefore, the fraction of LBGs that exhibit strong \lya emission lines can trace the fraction of neutral hydrogen in the IGM at $z\ge6$. This so-called \lya visibility test from previous works has shown that the fraction of LBGs with strong \lya emission increases steadily from $z \sim 2$ to 6 and then declines sharply towards $z \sim 7$ (e.g., \citealt{stark_keck_2010,stark_keck_2011,ono_spectroscopic_2012,schenker_keck_2012,finkelstein_galaxy_2013,tilvi_rapid_2014,vanzella_52_2014,cassata_vimos_2015,schmidt_grism_2016,jung_texas_2018,jung_clear_2022}). This is broadly consistent with the evolving neutral fraction in the IGM at the end of the reionization epoch.

In this paper, we present a sample of 45 LBGs with spectroscopically confirmed \lya emission lines at $z \sim 6$ from a survey of high-redshift galaxies using the Magellan telescope \citep{jiang_magellan_2017}. In Section 2, we introduce the survey program, target selection, spectroscopic observations, and data reduction. In Section 3, we construct our spectroscopic sample of LBGs. In Section 4, we present the properties of the galaxies using spectra and images from multi-wavelength observations. We discuss our results in Section 5, and summarize our paper in Section 6. Throughout the paper, we use a standard flat cosmology with $\rm H_0 = 70\ km\ s^{-1}\ Mpc^{-1}$, $\Omega_m = 0.3$ and $\Omega_\Lambda = 0.7$. All magnitudes refer to the AB system.

\section{Target Selection and Spectroscopic Observations} \label{sec:data}

\subsection{A Magellan M2FS Survey of $z\sim 6$ Galaxies} \label{subsec:survey}

\begin{deluxetable*}{cccccccccc}
\tablewidth{\textwidth} 
\tablenum{1}
\tablecaption{Survey Fields\label{fields}}

\tablehead{
\colhead{Fields} & \colhead{Coordinates} & \colhead{Area} & \colhead{Filters} & \colhead{$R/r'$} & \colhead{$I/i'$} & \colhead{$z'$} & \colhead{Candidates}   & \colhead{Good} & \colhead{Possible} \\
\colhead{} & \colhead{(J2000)} & \colhead{(deg$^2$)} & \colhead{} & \colhead{(mag)} & \colhead{(mag)} & \colhead{(mag)} & \colhead{} & \colhead{} & \colhead{} 
}
\decimalcolnumbers
\startdata
 SXDS & 02:18:00--05:00:00 &$ 1.12 $&$ R, i', z' $&$ 27.4 $&$ 27.4 $&$ 26.2 $&$ 368  $&$ 25 $&$ 7 $\\
 A370a & 02:39:55--01:35:24 &$ 0.16 $&$ R, I, z' $&$ 27.0 $&$ 26.2 $&$ 26.3 $&$ 86 $&$ 3 $&$ 2 $\\
 ECDFS & 03:32:25--27:48:18 &$ 0.22 $&$ r', i', z' $&$ 27.4 $&$ 27.5 $&$ 26.7 $&$ 105 $&$ 3 $&$ 2 $\\
 COSMOS & 10:00:29+02:12:21 &$ 1.26 $&$ r', i', z' $&$ 26.7 $&$ 26.3 $&$ 25.5 $&$ 145 $&$ 3 $&$ 0 $\\
\enddata
\tablecomments{Col.(1): Field names. Col.(2): Center coordinates. Col.(3): Area coverage. Col.(4): Broadband filters. Col.(5-7): Magnitude limits ($5\sigma$ detections in a $2''$-diameter aperture). Col.(8): Numbers of LBG candidates. Col.(9-10): Numbers of spectroscopically confirmed LBGs in the `good' sample and the `possible' sample.}
\end{deluxetable*}

We carried out a large spectroscopic survey of galaxies at $5.5 < z < 6.8$, using the large field-of-view (FoV), fiber-fed, multi-object spectrograph M2FS \citep{mateo_m2fs_2012} on the 6.5 m Magellan Clay telescope. The overview of the survey program was provided in \citet{jiang_magellan_2017}. The survey was designed to build a large and homogeneous sample of high-redshift galaxies, including LAEs at $z \approx 5.7$ and 6.6, and LBGs with strong \lya emission at $5.5 < z < 6.8$. Based on this sample, we can study the properties of these galaxies, the \lya luminosity function (LF) and its evolution at high redshift, high-redshift protoclusters, cosmic reionization, etc. Taking advantage of a $30'$-diameter FoV, M2FS is very efficient in identifying relatively bright high-redshift galaxies (e.g., \citealt{oyarzun_how_2016,oyarzun_comprehensive_2017}). The fields that we chose to observe are well-studied deep fields, including the Subaru XMM-Newton Deep Survey (SXDS), A370, the Extended Chandra Deep Field- South (ECDFS), COSMOS, etc. They cover a total of $\sim 2$~deg$^2$. These fields have deep Subaru Suprime-Cam imaging data in a series of broad bands $BV R(r')I(i')z'$ and narrow bands (e.g., NB816 and NB921). The images have been used to select LAEs at  $z \approx 5.7$ and 6.6 and LBGs at $z \sim 6$ in the survey program. The fields are summarized in Table~\ref{fields}. Some of the fields were later observed by Subaru Hyper Suprime-Cam (HSC) \citep{Aihara_2022_HSCSSP}, but we did not use the HSC data by the time of our observations.
 
The M2FS observations of the program have been completed and the data have been reduced. From the survey,  \citet{ning_magellan_2020} built a large sample of 260 LAEs at $z \approx 5.7$ and \citet{jiang_giant_2018} discovered a giant protocluster at $z \approx 5.7$. \citet{ning_magellan_2022} then assembled a sample of 36 LAEs at $z \approx 6.6$. By comparing with the $z \approx 5.7$ \lya LF, they confirmed that there is a rapid LF evolution at the faint end, but there is a lack of evolution at the bright end. \citet{wu_diffuse_2020} detected diffuse \lya halos around galaxies at $z \approx 5.7$ by stacking 310 spectroscopically confirmed LAEs. In this paper, we focus on luminous LBGs at $z\sim6$. The details of the target selection is presented in Section 2.3.

\subsection{Spectroscopic Observations and Data Reduction} \label{subsec:obs}

\begin{figure*}[t]
\centering
\includegraphics[width=\textwidth]{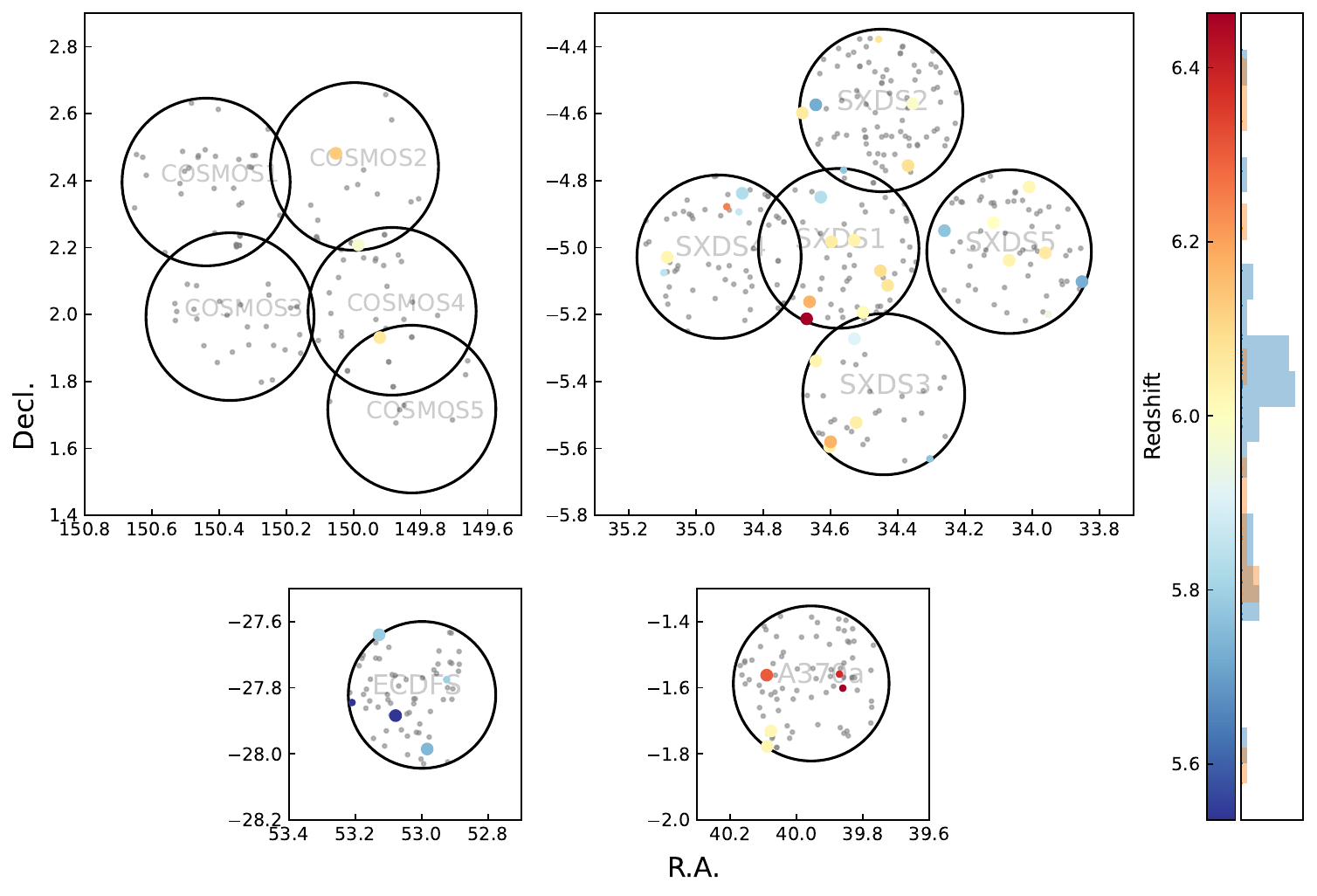}
\caption{Deep fields used in this work. The large circles represent the M2FS pointings. All points represent the LBG targets observed by our M2FS survey. The color-coded points represent the spectroscopically confirmed LBGs with \lya emission: the large points represent LBGs in the `good' sample and the small points represent LBGs in the `possible' sample. The colors indicate their spectroscopic redshifts measured from the \lya lines. The histogram on the right shows the the redshift distribution of the `good' sample (blue) and the `possible' sample (orange). \label{loc}}
\end{figure*}

We carried out the M2FS observations in 2015–2018. The resolving power of the spectra is $R \sim 2000$, and the wavelength coverage is roughly from 7600 to 9600~\AA, corresponding to a redshift range of $z = 5.3–6.8$ for the detection of the \lya emission line. The selection of M2FS pointing centers was limited by the number and spatial distribution of bright stars that were used for guiding, alignment, and primary-mirror wavefront corrections. The layout of the pointings is shown in Figure~\ref{loc}. The effective integration time per pointing was about 5~hr on average, and the individual exposure time was 30~min, 45~min, or 1~hr, depending on airmass and weather conditions. 

We used our own customized pipeline for data reduction, including bias (overscan) correction, dark subtraction, flat-fielding, cosmic ray identification, wavelength calibration, and sky subtraction. For detailed steps, see \citet{jiang_magellan_2017}. The pipeline produces combined 1D and 2D spectra as well as the 2D spectra of individual exposures. This pipeline is performed for all targets in the same fields, including the candidates of LAEs and LBGs. The final 1D and 2D spectra were used to identify \lya emission lines.

\subsection{Selection of $z\sim6$ LBG Candidates} \label{subsec:candidate}

\begin{figure}[t]
\centering
\includegraphics[width=0.48\textwidth]{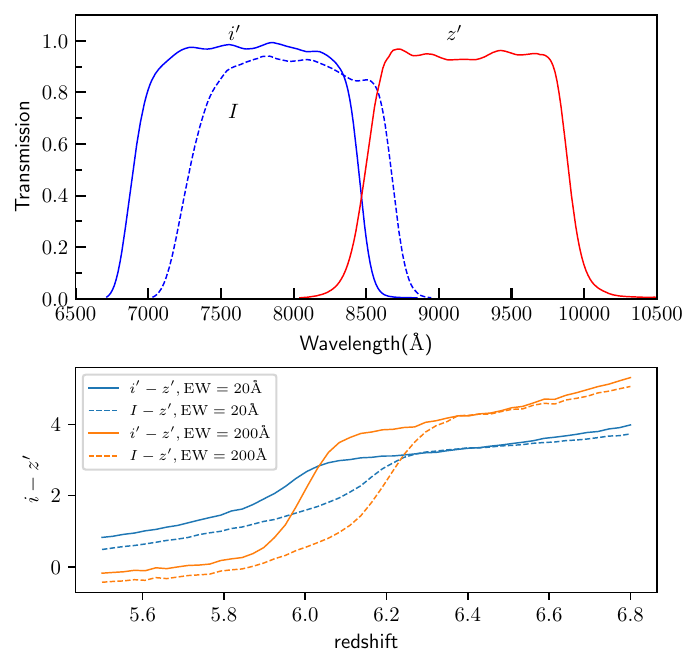}
\caption{Upper panel: transmission curves of the Suprime-Cam broadband filters that were used for our LBG candidate selection. Lower panel: expected values of $i’-z’$ or $I-z’$ as a function of redshift. The simulated LBG spectra have a UV continuum slope of --2.0 and a \lya emission line with two EW values, and the IGM absorption model from \citet{Inoue_2014_igm} is applied. \label{trans_curve}}
\end{figure}

Our M2FS observations included a sample of luminous LBG candidates at $z\sim6$. Deep Subaru Suprime-Cam images were used for the selection of these candidates. The selection was mainly based on the $i - z'$ color (here $i$ means either $i'$ or $I$). The upper panel of Figure~\ref{trans_curve} shows the filters that were used for our target selection. The selection limit is $7\sigma$ detections in the $z'$ band. We applied the following color cuts, 
\[i - z' > 1.3\ {\rm or}\ i < 3 \sigma\ {\rm detection},\]  
\[r - z' > 2.0\ {\rm or}\ r < 3 \sigma\ {\rm detection}.\]   

To illustrate the evolution of the $i - z'$ color with redshift, we produced a simple galaxy spectrum consisting of a power-law continuum and a \lya emission line. The power-law continuum has a fixed slope of --2.0, and the \lya equivalent width (EW) has two values, 20 and 200 \AA, respectively.  We then apply the \citet{Inoue_2014_igm} IGM attenuation models to the spectra.
The result is shown in the lower panel of Figure~\ref{trans_curve}. The value of $i - z'$ increases rapidly at $z \sim 6$. With our selection criteria, galaxies with $z \gtrsim 5.6$ can be selected. The actual selection function is more complex and depends on the galaxy continuum emission and \lya line emission. We also required no detections ($<2\sigma$) in any bands bluer than $r$ (or $R$), assuming that no flux can be detected at the wavelength bluer than the Lyman limit. Each candidate was visually inspected.

Compared to the previous selection criteria of LBGs  (e.g., \citealt{ono_great_2018_GOLDRUSH,bouwens_new_2021}), our selection criteria are relatively conservative. Previous studies usually applied redder $i-z$ colors (e.g., $i-z>1.5$ in  \citealt{ono_great_2018_GOLDRUSH}), and often considered infrared photometry. For example, many contaminants of high-redshift galaxies are late-type dwarf stars or low-redshift red (dusty) galaxies, and thus tend to have redder colors in the near-IR. We did not use near-IR data, despite the fact that some of our fields are covered by deep near-IR images. Our conservative criteria allow us to include less promising candidates and thus achieve higher sample completeness. On the other hand, it means a relatively lower efficiency, i.e., a larger fraction of contaminants. It is not a concern in our program, since we have enough fibers to cover almost all these candidates. In Table~\ref{fields}, Column 8 shows the numbers of the candidates in each field that were observed by our program.


\section{Spectroscopic Results} \label{sec:result}

\subsection{\lya Line Identification and a Sample of $z\sim6$ LBGs} \label{subsec:identification}

In this subsection, we will describe how to identify \lya emission lines in the spectra, and present two samples of spectroscopically confirmed LBGs, including a `good' sample with high confidence and a `possible' sample with lower confidence. We will then derive some basic properties of the \lya lines.

We use both 1D and 2D spectra to identify \lya emission lines. We first carry out an initial search. For each galaxy, we bin its 1D spectrum by 3 \AA\ and search for an emission line with signal-to-noise ratio S/N $> 5$ in the whole wavelength range. For a line in the `good' sample, it needs to cover at least 5 contiguous pixels with S/N $> 1$ in each pixel in the binned spectrum. This is determined by the typical  \lya emission line width in our spectra. For the `possible' sample, we allow a line to cover only 3 or 4 contiguous pixels with S/N $> 1$. The S/N of a line is estimated by summing up the corresponding pixels of the line in the original spectrum. The reason of adopting a criterion of S/N $> 1$ is that most of the regions in the LBG spectra are contaminated by strong OH sky lines. If a \lya line is close to OH lines, it could be severely affected by sky line residuals after the sky background subtraction.

After the initial screening above, we visually inspect each identified emission line in the 2D spectra, including the combined spectrum and the spectra of individual exposures. Obvious false detections are removed. The \lya emission line of a $z\sim6$ galaxy is often much broader than other emission lines at much lower redshift  in the observed frame due to the (1+$z$) broadening. It also shows an asymmetric profile due to strong IGM absorption and ISM kinematics. Possible contaminant lines from low-redshift galaxies usually appear narrow and symmetric, including H$\beta$, $[$\ion{O}{3}$]$ $\lambda$5007, or H$\alpha$ emission lines. In addition, our resolving power of $\sim 2000$ can nearly resolve the $[$\ion{O}{2}$]$ $\lambda \lambda$3727, 3729 doublet. 
To quantify the asymmetry of the \lya line, we calculate the weighted skewness $S_W$ introduced by \citet{Shimasaku_2006_skewness}. This is the skewness (or third moment) of the line multiplied by $(\lambda_{10,r} - \lambda_{10,b})$, where $\lambda_{10,r}$ and $\lambda_{10,b}$ are the wavelengths where the flux drops to $10\%$ of its peak value at the red and blue sides of the line, respectively. Since the \lya emission of high-redshift galaxies tends to be broader than other emission lines of lower-redshift galaxies, this line width factor enhances the difference between \lya and other lines. 
The \lya $S_W$ values of the LBGs in our `good' sample and `possible' sample range from 2.3 to 18.4 and from 2.1 to 12.2, respectively. In comparison, we also calculate $S_W$ of 10 emission lines that are identified as low-redshift interlopers based on their double-peak (e.g., an $[$\ion{O}{2}$]$ $\lambda \lambda$3727, 3729 doublet at $z\sim1.3$), or based on the detection of other lines at the same redshift in the spectra (e.g., an $[$\ion{O}{3}$]$ $\lambda$5007 and corresponding $[$\ion{O}{3}$]$ $\lambda$4959 at $z\sim0.72$). Their $S_W$ values span a range from $-12.5$ to $3.3$.
Usually $S_W>3$ has been used as a threshold to distinguish \lya from other lines, but the asymmetric shape of a high-redshift \lya emission line may not be obvious when its S/N is relatively low \citep[e.g.,][]{ning_magellan_2022}. For three LBGs with $2.1<S_W<3$, they were not detected in two blue bands $B$ and $V$, so we put them in the `possible' sample.

We emphasize that in our LBG candidate selection procedure addressed in Section 2.3, we required non-detections for the candidates in the $B$- and $V$-band images. These images reach a great depth of $\ge28.5$ mag. They can efficiently exclude low-redshift interlopers.

\begin{figure*}[t]
\centering
\includegraphics[width=\textwidth]{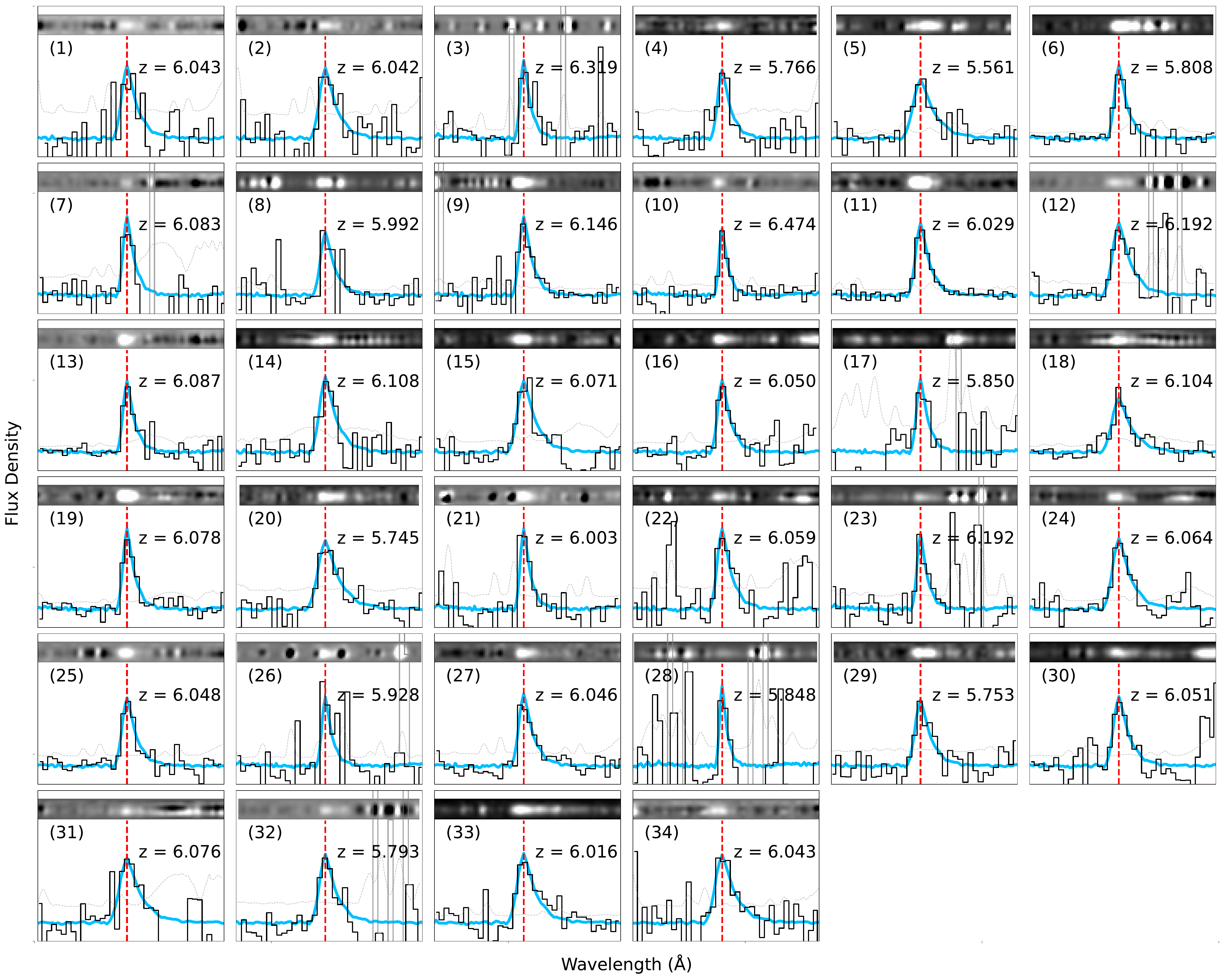}
\caption{M2FS 1D and 2D spectra of the 34 LBGs in our $z \approx 6.6$ sample. In each panel, the 1D spectrum has been binned to 3~\AA\ per pixel in arbitrary units. The light grey line indicates the $1\sigma$ uncertainty. The best-fit \lya line template is over-plotted in blue. The red dashed line shows the peak position of the \lya emission line. The source numbers correspond to those shown in Column 1 in Table 2. \label{1D2Dspec}}
\end{figure*}

Finally, the `good' sample of 34 LBGs has convincing, luminous \lya emission lines with S/N $> 5$ and asymmetric profiles. Figure~\ref{1D2Dspec} shows the 1D and 2D spectra in the \lya region of the 36 LBGs in the `good' sample. We can see that strong emission lines clearly show asymmetric line shapes due to the IGM absorption and ISM kinematics. The `possible' sample contains 11 LBGs with less promising \lya emission lines. We show all the confirmed LBGs in Table~\ref{sample}. Column 5 lists the spectroscopic redshifts measured from the \lya lines, and their errors are smaller than 0.001. Figure~\ref{loc} illustrates the positions of the targets in the five fields, including the observed candidates (all points) and the spectroscopically confirmed LBGs (color-coded points). The large circles represent the M2FS pointings. Despite the fact that the exposure time and depth of the individual pointings are similar, the numbers of the confirmed LBGs in these pointings are quite different, suggesting the existence of significant cosmic variance. As mentioned in \citet{ning_magellan_2020,ning_magellan_2022}, COSMOS1 and COSMOS3 have some alignment problems during the observations, so they only have a few LBGs confirmed.

\subsection{Properties of the \lya Emission Lines} \label{subsec:properties}

We measure redshifts and \lya line flux for the LBGs using a \lya profile template. The template is a composite \lya spectrum of a large number of LAEs at $z \approx 5.7$ \citep{ning_magellan_2020}. For each LBG, we first estimate an initial redshift using the wavelength of the \lya line peak (1215.67~\AA). We then refine this redshift and model the \lya profile as follows. From the \lya template, we generate a set of model spectra for a grid of peak values, line widths, and redshifts. The peak values vary from 0.9 to 1.1 (times of the observed peak value) with a step size of 0.01. The line widths are from 0.5 to 2.0 times the original width with a step size of 0.1. The redshift values vary within the initial redshift $\pm 0.002$ with a step size of 0.0001. We find the best-fit model for each LBG using the $\chi^2$ method. The best-fit redshifts are listed in Column~5 of Table~\ref{sample}.

\begin{deluxetable*}{ccccccccccc}
\tabletypesize{\scriptsize}
\tablewidth{\textwidth} 
\tablenum{2}
\tablecaption{Confirmed LBGs with \lya emission lines \label{sample}}

\tablehead{
\colhead{No.} & \colhead{Field} & \colhead{R.A.} & \colhead{Decl.} & \colhead{Redshift} & \colhead{$i$} & \colhead{$z'$} & \colhead{log$_{10}$L(Ly$\alpha$)}  & \colhead{$M_{\rm UV}$} &  \colhead{\lya EW} &  \colhead{$\beta_{\mathrm{UV}}$} \\
\colhead{} & \colhead{} & \colhead{(J2000)} & \colhead{(J2000)} & \colhead{} & \colhead{(mag)} & \colhead{(mag)} & \colhead{(erg/s)} & \colhead{(mag)}  &  \colhead{(\AA)}  & \colhead{}
}
\decimalcolnumbers
\startdata
\multicolumn{11}{c}{Good Sample}   \\ \hline
$ 1 $& A370a & 02:40:21.24 & --01:46:37.52 &$ 6.043 $&$ >27.2 $&$ 25.43 $&$ 42.62 $&$ -21.31 $&$ 14.77 $&$ - $\\
$ 2 $& A370a & 02:40:18.51 & --01:43:51.66 &$ 6.042 $&$ >27.2 $&$ 24.84 $&$ 42.73 $&$ -21.93 $&$ 10.66 $&$ - $\\
$ 3 $& A370a & 02:40:21.60 & --01:33:42.16 &$ 6.319 $&$ >27.2 $&$ 25.24 $&$ 43.10 $&$ -21.83 $&$ 27.14 $&$ - $\\
$ 4 $& ECDFS & 03:31:56.08 & --27:59:06.50 &$ 5.766 $&$ >28.5 $&$ 25.61 $&$ 42.58 $&$ -20.86 $&$ 39.40 $&$ 1.38 $\\
$ 5 $& ECDFS & 03:32:18.92 & --27:53:02.81 &$ 5.561 $&$ >28.5 $&$ 24.48 $&$ 43.37 $&$ -21.96 $&$ 44.83 $&$ -1.84 $\\
$ 6 $& ECDFS & 03:32:30.85 & --27:38:22.20 &$ 5.808 $&$ 27.84 $&$ 24.94 $&$ 43.26 $&$ -22.25 $&$ 44.38 $&$ 0.55 $\\
$ 7 $& COSMOS & 09:59:41.18 & +01:55:51.75 &$ 6.083 $&$ >27.3 $&$ 24.78 $&$ 42.65 $&$ -21.22 $&$ 14.87 $&$ -2.44 $\\
$ 8 $& COSMOS & 09:59:56.54 & +02:12:27.15 &$ 5.992 $&$ 27.12 $&$ 24.76 $&$ 43.27 $&$ -21.62 $&$ 42.13 $&$ -2.52 $\\
$ 9 $& COSMOS & 10:00:12.59 & +02:28:52.08 &$ 6.146 $&$ >27.3 $&$ 25.03 $&$ 43.42 $&$ -21.36 $&$ 87.75 $&$ - $\\
$ 10 $& SXDS1 & 02:18:41.01 & --05:12:47.18 &$ 6.474 $&$ 27.9 $&$ 25.72 $&$ 42.93 $&$ -21.76 $&$ 19.85 $&$ - $\\
$ 11 $& SXDS1 & 02:18:00.91 & --05:11:37.77 &$ 6.029 $&$ 27.67 $&$ 24.75 $&$ 43.35 $&$ -21.48 $&$ 52.53 $&$ -2.98 $\\
$ 12 $& SXDS1 & 02:18:38.92 & --05:09:44.02 &$ 6.192 $&$ >28.4 $&$ 25.06 $&$ 43.17 $&$ -21.55 $&$ 41.04 $&$ -1.93 $\\
$ 13 $& SXDS1 & 02:17:43.25 & --05:06:47.60 &$ 6.087 $&$ >28.4 $&$ 25.90 $&$ 43.25 $&$ -20.08 $&$ 189.33 $&$ -1.98 $\\
$ 14 $& SXDS1 & 02:17:48.40 & --05:04:10.79 &$ 6.108 $&$ >28.4 $&$ 25.67 $&$ 43.25 $&$ -19.80 $&$ 229.40 $&$ -2.22 $\\
$ 15 $& SXDS1 & 02:18:23.34 & --04:58:58.18 &$ 6.071 $&$ >28.4 $&$ 25.70 $&$ 43.20 $&$ -20.95 $&$ 53.24 $&$ -3.64 $\\
$ 16 $& SXDS1 & 02:18:07.14 & --04:58:41.50 &$ 6.050 $&$ >28.4 $&$ 24.84 $&$ 43.10 $&$ -21.61 $&$ 22.74 $&$ -3.64 $\\
$ 17 $& SXDS1 & 02:18:30.93 & --04:50:58.73 &$ 5.850 $&$ 27.04 $&$ 24.65 $&$ 42.67 $&$ -21.27 $&$ 13.86 $&$ -2.77 $\\
$ 18 $& SXDS2 & 02:17:28.71 & --04:45:19.61 &$ 6.104 $&$ >28.4 $&$ 25.67 $&$ 43.45 $&$ -19.48 $&$ 527.92 $&$ - $\\
$ 19 $& SXDS2 & 02:18:44.06 & --04:35:54.24 &$ 6.078 $&$ >28.4 $&$ 25.39 $&$ 43.26 $&$ -20.98 $&$ 85.69 $&$ - $\\
$ 20 $& SXDS2 & 02:18:34.48 & --04:34:25.86 &$ 5.745 $&$ 26.59 $&$ 25.19 $&$ 42.92 $&$ -22.21 $&$ 17.81 $&$ -0.21 $\\
$ 21 $& SXDS2 & 02:17:25.27 & --04:34:12.85 &$ 6.003 $&$ >28.4 $&$ 25.69 $&$ 42.98 $&$ -20.87 $&$ 49.57 $&$ - $\\
$ 22 $& SXDS3 & 02:18:24.72 & --05:35:41.82 &$ 6.059 $&$ 27.55 $&$ 25.67 $&$ 42.84 $&$ -20.97 $&$ 33.18 $&$ - $\\
$ 23 $& SXDS3 & 02:18:23.90 & --05:34:49.34 &$ 6.192 $&$ 28.19 $&$ 24.95 $&$ 42.94 $&$ -22.06 $&$ 24.16 $&$ 0.32 $\\
$ 24 $& SXDS3 & 02:18:05.68 & --05:31:21.90 &$ 6.064 $&$ >28.4 $&$ 25.55 $&$ 43.29 $&$ -20.65 $&$ 125.98 $&$ - $\\
$ 25 $& SXDS3 & 02:18:34.47 & --05:20:20.96 &$ 6.048 $&$ >28.4 $&$ 25.41 $&$ 43.05 $&$ -21.16 $&$ 44.85 $&$ - $\\
$ 26 $& SXDS3 & 02:18:06.96 & --05:16:21.93 &$ 5.928 $&$ 27.07 $&$ 25.26 $&$ 43.04 $&$ -20.95 $&$ 55.84 $&$ -1.64 $\\
$ 27 $& SXDS4 & 02:20:20.49 & --05:01:45.06 &$ 6.046 $&$ >28.4 $&$ 25.56 $&$ 43.02 $&$ -20.99 $&$ 49.65 $&$ - $\\
$ 28 $& SXDS4 & 02:19:26.97 & --04:50:17.71 &$ 5.848 $&$ 26.67 $&$ 25.27 $&$ 42.67 $&$ -21.40 $&$ 15.23 $&$ - $\\
$ 29 $& SXDS5 & 02:15:24.62 & --05:06:07.27 &$ 5.753 $&$ 26.34 $&$ 25.07 $&$ 42.86 $&$ -21.57 $&$ 19.90 $&$ - $\\
$ 30 $& SXDS5 & 02:16:16.53 & --05:02:17.80 &$ 6.051 $&$ >28.4 $&$ 25.51 $&$ 43.15 $&$ -20.88 $&$ 73.98 $&$ -1.82 $\\
$ 31 $& SXDS5 & 02:15:50.69 & --05:00:57.88 &$ 6.076 $&$ 27.32 $&$ 25.50 $&$ 43.06 $&$ -21.62 $&$ 48.69 $&$ 0.42 $\\
$ 32 $& SXDS5 & 02:17:02.70 & --04:56:59.28 &$ 5.793 $&$ 26.05 $&$ 24.28 $&$ 42.85 $&$ -21.96 $&$ 13.90 $&$ -1.73 $\\
$ 33 $& SXDS5 & 02:16:27.81 & --04:55:34.25 &$ 6.016 $&$ 27.36 $&$ 24.35 $&$ 43.37 $&$ -21.13 $&$ 77.92 $&$ -2.89 $\\
$ 34 $& SXDS5 & 02:16:02.36 & --04:49:08.81 &$ 6.043 $&$ >28.4 $&$ 25.60 $&$ 43.03 $&$ -20.94 $&$ 52.51 $&$ - $\\ \hline
\multicolumn{11}{c}{Possible Sample}   \\ \hline
$ 1 $& A370a & 02:39:26.57 & --01:36:03.84 &$ 6.463 $&$ >27.2 $&$ 25.59 $&$ 42.43 $&$ -22.05 $&$ 4.76 $&$ - $\\
$ 2 $& A370a & 02:39:28.99 & --01:33:31.00 &$ 6.383 $&$ 26.73 $&$ 25.84 $&$ 42.68 $&$ -21.51 $&$ 13.99 $&$ - $\\
$ 3 $& ECDFS & 03:32:50.43 & --27:50:40.23 &$ 5.536 $&$ 28.45 $&$ 25.63 $&$ 42.89 $&$ -21.16 $&$ 55.39 $&$ 0.85 $\\
$ 4 $& ECDFS & 03:31:41.99 & --27:46:30.91 &$ 5.795 $&$ >28.5 $&$ 25.56 $&$ 42.63 $&$ -21.09 $&$ 18.11 $&$ - $\\
$ 5 $& SXDS2 & 02:18:14.62 & --04:46:07.52 &$ 5.778 $&$ 27.5 $&$ 26.03 $&$ 42.47 $&$ -20.62 $&$ 19.60 $&$ - $\\
$ 6 $& SXDS2 & 02:17:49.61 & --04:22:41.36 &$ 6.074 $&$ >28.4 $&$ 25.74 $&$ 43.10 $&$ -20.65 $&$ 81.85 $&$ - $\\
$ 7 $& SXDS3 & 02:17:13.11 & --05:37:53.43 &$ 5.768 $&$ 26.32 $&$ 24.76 $&$ 42.64 $&$ -21.88 $&$ 9.04 $&$ - $\\
$ 8 $& SXDS4 & 02:20:22.93 & --05:04:31.83 &$ 5.849 $&$ 25.81 $&$ 24.23 $&$ 43.08 $&$ -22.74 $&$ 18.18 $&$ 0.40 $\\
$ 9 $& SXDS4 & 02:19:29.35 & --04:53:38.27 &$ 5.862 $&$ 27.83 $&$ 25.38 $&$ 42.47 $&$ -21.69 $&$ 13.55 $&$ 1.11 $\\
$ 10 $& SXDS4 & 02:19:37.99 & --04:52:41.57 &$ 6.246 $&$ 28.14 $&$ 25.65 $&$ 43.10 $&$ -20.98 $&$ 52.87 $&$ -2.41 $\\
$ 11 $& SXDS5 & 02:15:48.66 & --05:11:58.28 &$ 5.938 $&$ 27.32 $&$ 25.02 $&$ 43.45 $&$ -21.53 $&$ 79.60 $&$ - $\\\enddata
\tablecomments{Col.(1): Object numbers. Col.(2): Field names. Col.(3-4): Source coordinates. Col.(5): Spectroscopic redshifts (their errors are smaller than 0.001). Col.(6-7): Subaru Suprime-Cam $i$ band and $z'$ band magnitudes. The upper limits indicate $2\sigma$ detections. Col.(8): \lya line luminosities. Col.(9): Absolute UV magnitudes at rest-frame 1500 \AA. Col.(10): \lya line EWs.  Col.(11): UV slopes calculated from multi-band data.}
\end{deluxetable*}

\begin{figure}[!t]
\centering
\includegraphics[width=0.48\textwidth]{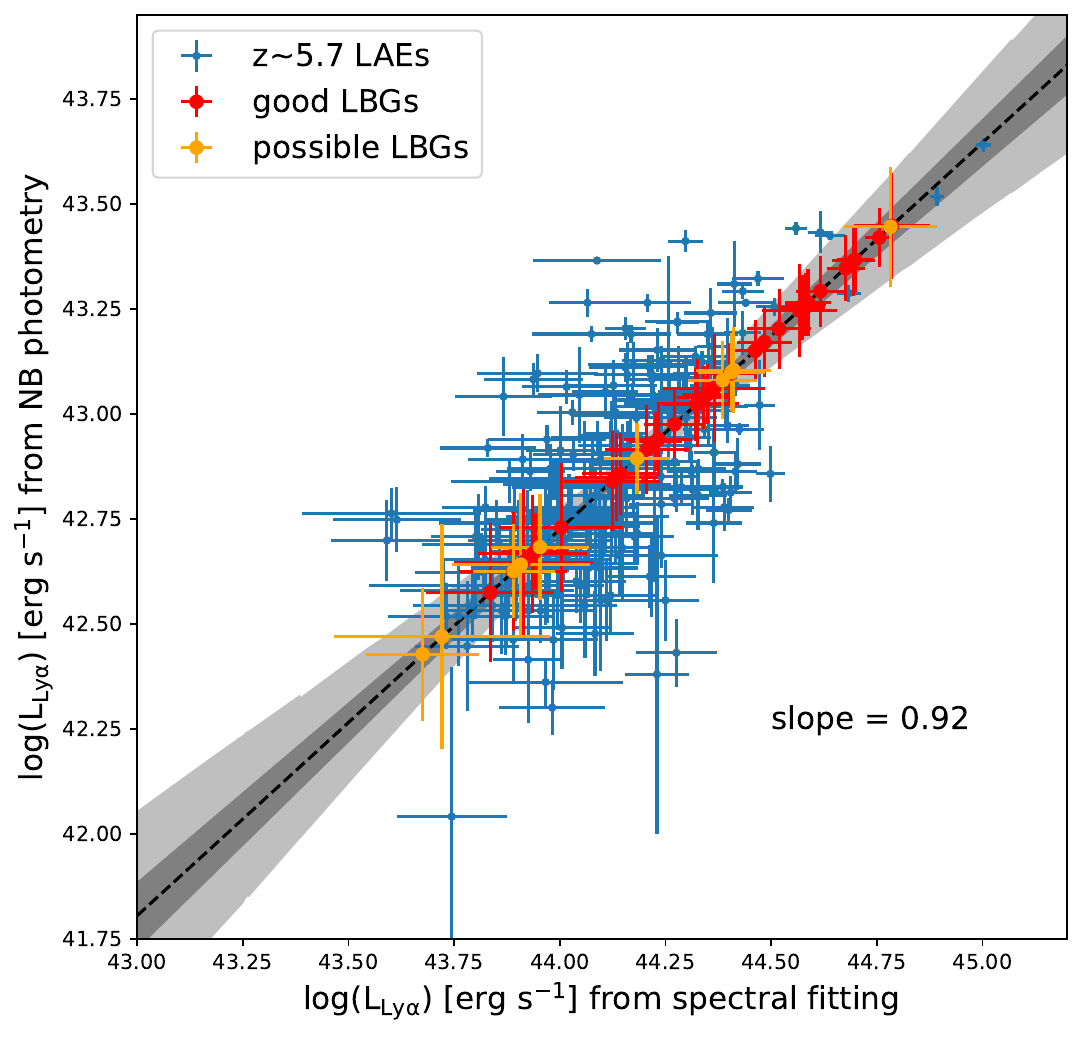}
\caption{Relation between the \lya luminosities L(Ly$\alpha$) calculated from the spectral fitting method and from the photometric data (see details in Section 3.2). The blue points represent the $z \approx 5.7$ LAE sample in \citet{ning_magellan_2020} and the red points represent our LBG sample. The best-fit relation from the $z \approx 5.7$ LAE sample is depicted by the black dashed line, and the shaded regions indicate its $1\sigma$ and $3\sigma$ confidence regions, respectively. We apply this relation to estimate the \lya luminosities of the LBGs in this work. \label{Llya}}
\end{figure}

We measure the \lya line flux of each LBG by integrating the best-fit \lya model profile and applying a correction. Line flux directly calculated from an observed line in faint targets (like our LBGs) can be significantly underestimated due to several reasons, such as fiber loss, alignment problems, skyline contamination, etc. For narrowband-selected LAEs, their \lya line flux can be well determined using the combination of spectroscopic redshifts, narrowband photometry, and broadband photometry. This has been applied to the LAEs in our fields \citep{ning_magellan_2020,ning_magellan_2022}. For 260 LAEs at $z\approx 5.7$ in our fields, we compare their \lya flux derived from the photometric data and from the best-fit \lya line profiles, and the result is shown in Figure~\ref{Llya}. 
We use a hierarchical Bayesian method to do a linear regression on the \lya luminosities calculated by these two methods, considering the errors on both axes. The best-fit function $y=0.92x + 2.32$ is depicted by the black dashed line in Figure~\ref{Llya}, and the $1\sigma$ and $3\sigma$ confidence intervals are shown in the shaded regions.
Given that these LAEs and the LBGs in this work are in the same fields and were observed at the same time, the LBGs in our sample should follow the same relation. We apply this relation to measure the line flux of our LBGs, and the measured \lya luminosities L(Ly$\alpha$) are listed in Column~8 of Table~\ref{sample}. The IGM absorption blueward of \lya is not considered.

We use LBGs in the `good' sample to study the dependence of the \lya line width (full width at half maximum, or FWHM) on redshift and \lya luminosity. We exclude three galaxies whose \lya lines are likely largely affected by strong residuals of the sky subtraction. We then divide the remaining galaxies into two \lya luminosity bins, and produce a stacked spectrum for each bin. We adjust the number of spectra in each bin so that the two stacked spectra have similar S/N~$\sim 50$. The result is shown in the upper panel of Figure~\ref{FWHM}. We also produce two stacked spectra for two redshift bins using the same method, and the result is shown in the lower panel of Figure~\ref{FWHM}. The figure does not show apparent dependence of the line FWHM on redshift or luminosity. 
Using a large sample of LAEs at $z \approx 5.7$ and 6.6, \citet{ning_magellan_2020,ning_magellan_2022} found that more luminous galaxies tend to have higher \lya FWHMs, because more massive host halos possess higher HI column densities and higher gas velocities. We do not see this trend in our LBG sample, likely because the line shape has been severely affected by OH skylines. Unlike LAEs at $z \approx 5.7$ and 6.6, the \lya lines of the LBGs in our sample are often contaminated by skylines.

\section{Properties of the LBGs} \label{sec:result}

\subsection{UV slopes and \lya EWs} \label{sec:slope}

\begin{figure}[!t]
\centering
\includegraphics[width=0.48\textwidth]{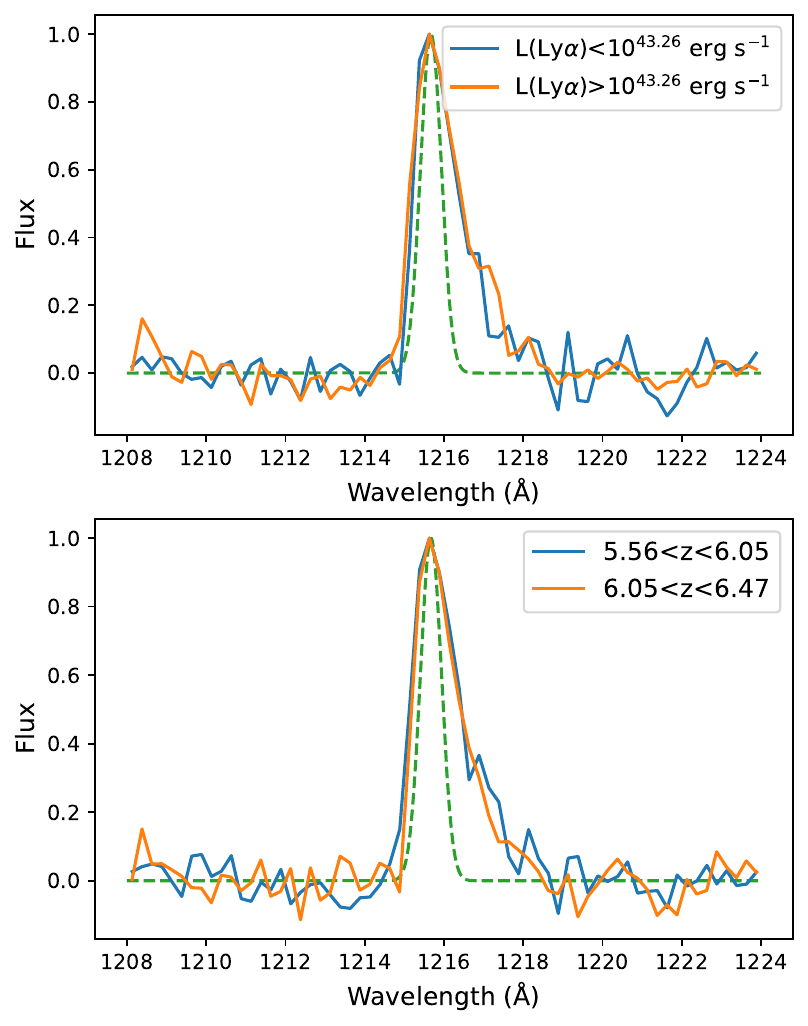}
\caption{Stacked \lya line spectra for the LBGs in two \lya luminosity bins (upper panel) and two redshift bins (lower panel). The spectra have been normalized so that the peak values are 1. The dashed lines represent the instrument resolution assuming a Gaussian profile. \label{FWHM}}
\end{figure}

In this section, we collect multi-wavelength observations to study the physical properties and stellar populations of the LBGs in our sample. We make use of the following data, including the Hyper Suprime-Cam Subaru Strategic Program (HSC-SSP, \citealt{Aihara_2022_HSCSSP}), the UKIRT Infrared Deep Sky Survey (UKIDSS, \citealt{Warren_2007_UKIDSS}), the Great Observatories Origins Deep Survey (GOODS, \citealt{Giavalisco_2004_goods}), the VISTA Deep Extragalactic Observations Survey (VIDEO, \citealt{Jarvis_2013_vista}), the Cosmic Assembly Near-IR Deep Extragalactic Legacy Survey (CANDELS, \citealt{Grogin_2011_candles,Koekemoer_2011_candles}),  and the Cosmic Dawn Survey by Spitzer Space Telescope (only the 3.6, 4.5~$\mathrm{\mu m}$ data are used, \citealt{Euclid_2022_irac}).  
The No.~13 LBG is covered by the JWST Cycle-1 program (GO 1837), Public Release IMaging for Extragalactic Research (PRIMER, \citealt{Dunlop_2021_primer}), and has NIRCam imaging in eight bands (F090W, F115W, F150W, F200W, F277W, F356W, F410M, and F444W). \citet{Ning_2022_JWST} carried out a pilot study of seven spectroscopically confirmed $z \sim 6$ LBGs, including No.~13 LBG,  using the JWST NIRCam imaging data. The No.~7 LBG is covered by another JWST Cycle-1 program COSMOS-Web (GO 1727, \citealt{Caitlin_2022_cosmosweb}), and has NIRCam imaging in four bands (F115W, F150W, F277W, and F444W).

Most LBGs in our sample except those in the A370a field have multi-wavelength data that can be used to determine UV continuum slope $\beta_{\mathrm{UV}}$ and other physical properties via SED modeling. 
The IRAC images of many LBGs are blended with nearby sources. In this case, GALFIT \citep{galfit:2002,galfit:2010} is used to model and subtract contaminant sources. Photometry of the target is then done after the de-blending. 

We then measure $\beta_{\mathrm{UV}}$ and the absolute UV magnitude $M_{\rm UV}$ (at rest-frame 1500 \AA) for the LBGs using the near-IR data mentioned above. The  median $\beta_{\mathrm{UV}}$ value is $-1.86$, which is consistent  with the median $\beta_{\mathrm{UV}} = -1.73^{+0.14}_{-0.20}$ for $z\sim 6$ LBGs within the absolute magnitude bin $M_{\rm UV} = -20.75 \pm 0.5$~mag from \citet{bouwens_uv-continuum_2014}. We also measure the UV slopes of bright LAEs at $z\approx 5.7$ in \citet{ning_magellan_2020} and find a median value of $\beta_{\mathrm{UV}}\sim-2$, which is consistent with the value for the LBGs in this sample. For the LBGs in A370a that do not have near-IR data, we use the $z'$ band photometry to estimate the UV continuum flux. Since we are not able to determine $\beta_{\mathrm{UV}}$, we adopt the median $\beta_{\mathrm{UV}} = -1.86$ from the other fields.
For simplicity, we assume that the flux blueward of \lya is completely absorbed. If a \lya line is in the $z'$ band, its flux is subtracted from the $z'$-band photometry. The calculated $M_{\rm UV}$ and \lya rest-frame EW values are listed in Columns 9 and 10 of Table~\ref{sample}. 

\begin{figure}[!t]
\centering
\includegraphics[width=0.5\textwidth]{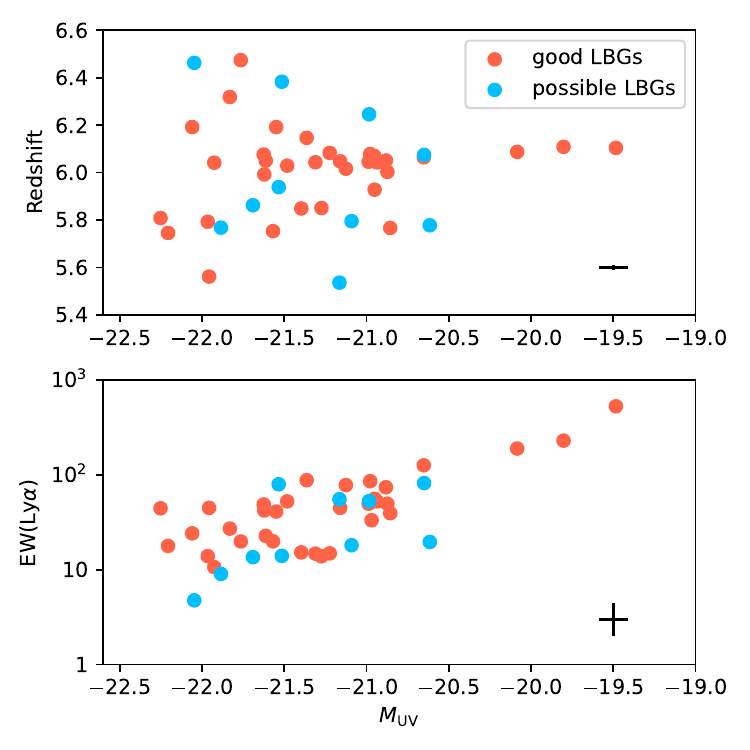}
\caption{ Distributions of the LBG redshifts (upper panel) and \lya EWs (lower panel) with respect to $M_{\rm UV}$. The `good' LBG sample and `possible' LBG sample are both presented. The median errors are shown in the lower right corner, and the error bar of redshift is too small to be seen. \label{M1500_z_EW}}
\end{figure}

Figure~\ref{M1500_z_EW} shows the distribution of the LBG redshifts and \lya EWs with respect to $M_{\rm UV}$ in our sample. These LBGs span a UV magnitude range of $-22.5$ to $-19.0$~mag, corresponding to $0.3-3.4$ times the characteristic luminosity of galaxies at $z\sim 6$ \citep{bouwens_new_2021}, and a \lya luminosity range of $(0.4-3) \times 10^{42}$ erg s$^{-1}$. They represent the most luminous galaxies at $z \sim 6$ in terms of both UV continuum luminosity and \lya luminosity.

\subsection{Stellar populations} \label{sec:SED}

\begin{figure}[!t]
\centering
\includegraphics[width=0.4\textwidth]{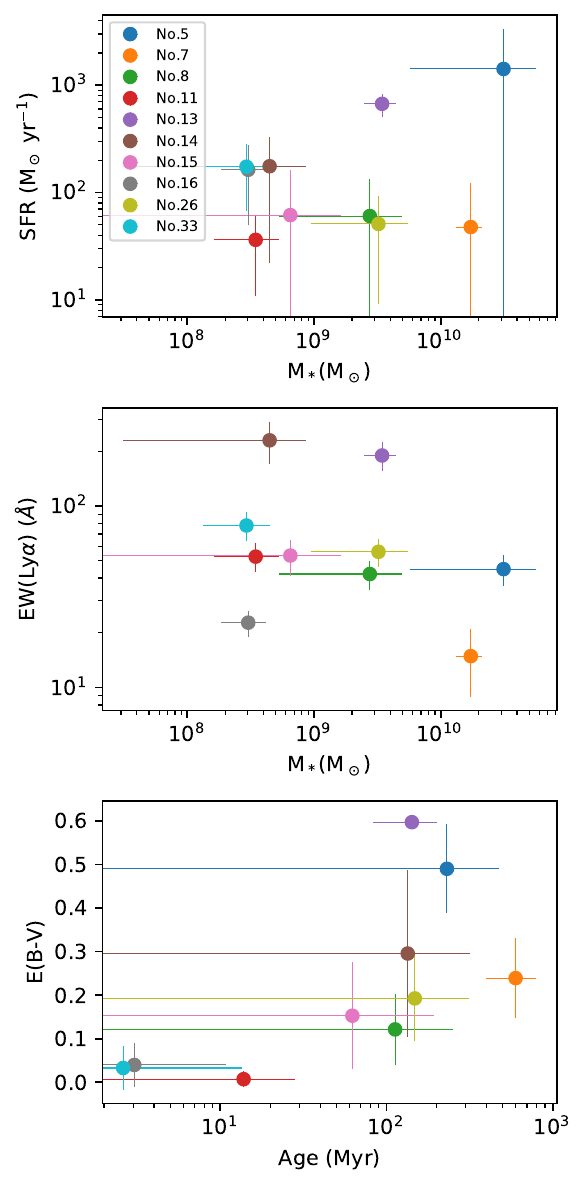}
\caption{ Properties of the ten LBGs measured from the SED modeling. They all cover wide ranges. \label{properties}}
\end{figure}

Ten LBGs in the `good' sample have sufficient multi-wavelength imaging data (especially IRAC $3.6~\mu m$ and $4.5~\mu m$ photometry) that allow us to perform SED modeling. We model their SEDs and derive their stellar populations using the Code Investigating GALaxy Emission  (CIGALE; \citealt{cigale:2009}). Our spectroscopic redshifts remove one critical free parameter in the fitting process. 
We use the stellar population synthesis models of \citet{Bruzual_2003_bc03} and adopt the Salpeter initial mass function. Given the limited number of available photometric data points, we use as few free parameters as possible. We use constant SFRs and fix metallicity to be 0.2 $Z_\sun$. A nebular emission template based on \citet{Inoue_2011_nebular} is included, with ionization parameter $U$, metallicity, and line width fixed. We do not include AGN components during the model fitting because these LBGs do not show any AGN features in the observed bands. Furthermore, \lya emission and radiative transfer in galaxies are very complex (e.g., \citealt{Stephane_1993_lya}). Our LBGs are among galaxies with the strongest \lya emission, and the nebular emission model built in CIGALE can not reproduce the \lya emission in the SEDs of many LBGs unless we assume an extremely young stellar population (see Section~\ref{sec:SED_lyaNB}). Therefore, we mask out the \lya line in the nebular template of CIGALE when fitting and subtracting the \lya contribution from the broadband photometry. 

\begin{figure*}[!t]
\centering
\includegraphics[width=0.9\textwidth]{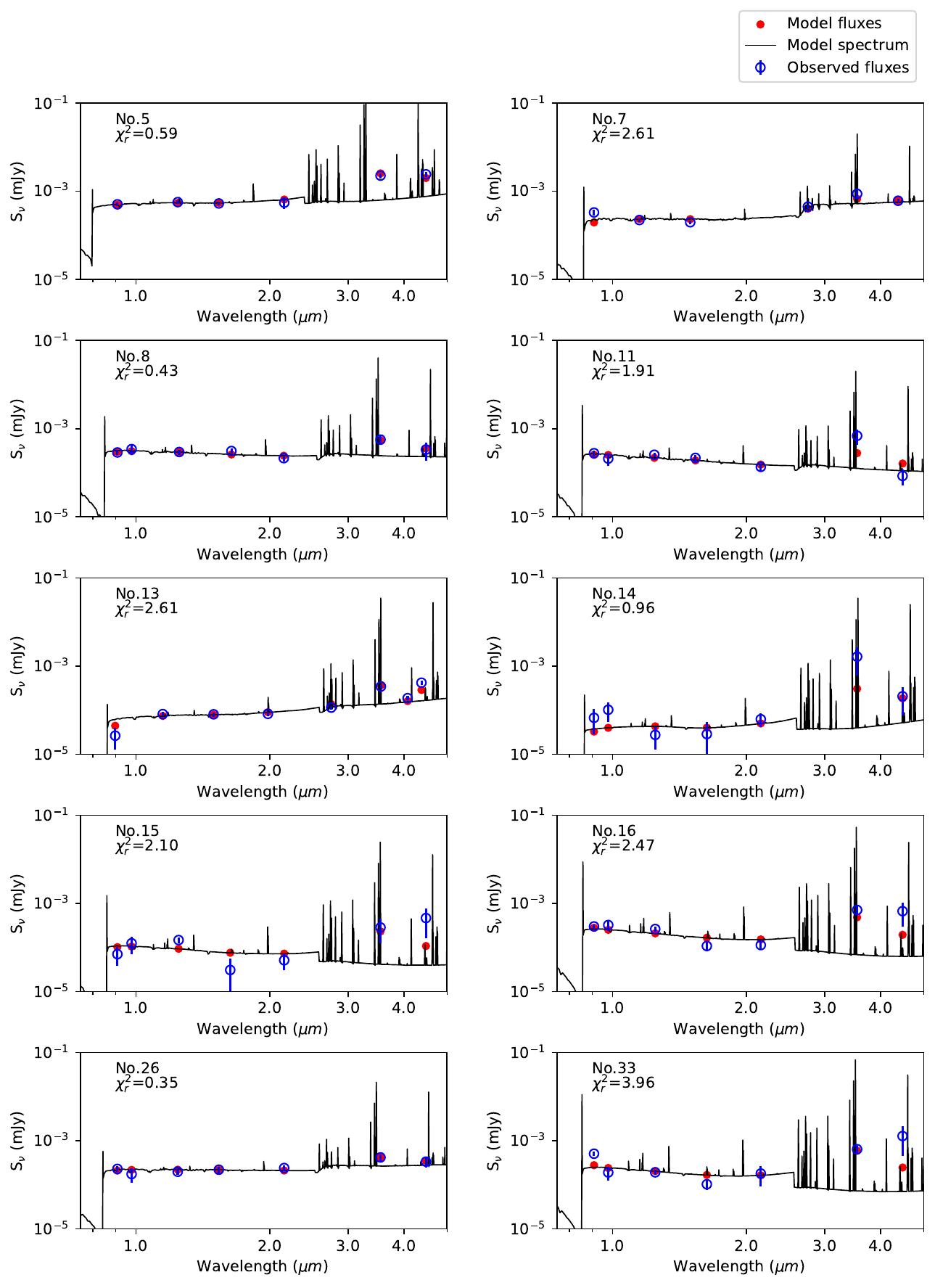}
\caption{ Results of the SED modeling for the ten LBGs in our sample. Upper panel: SFR versus stellar mass, both from SED modeling. Middle panel: \lya line EW from spectra versus stellar mass from SED modeling. Lower panel: dust reddening $E(B-V)$ versus stellar age of the main stellar population. The blue circles represent the observed fluxes and the red points indicate the best-fit model fluxes in the corresponding bands. \label{seds}}
\end{figure*}

We mainly constrain stellar mass $M_*$, age, star formation rate (SFR), and dust reddening $E(B-V)$. We use the dust extinction law of \citet{Calzetti_2000_dust} with $E(B-V)$ in a range of $0-0.6$. We allow age to vary between 1 and 800 Myr (the age of the Universe at $z=6$ is 914 Myr). The distributions of the derived properties are shown in Figure~\ref{properties} and the SED modeling results are shown in Figure~\ref{seds}. 
As seen from the figures, the derived values of the properties all span wide ranges, indicating that these LBGs have a variety of stellar populations. The LBGs are strongly star-forming galaxies with SFRs roughly between 30 and 1000 $M_\odot$ yr$^{-1}$. While most of them have SFRs around 100 $M_\odot$ yr$^{-1}$, two LBGs, (No.~5 and No.~13) have SFRs close to 1000 $M_\odot$ yr$^{-1}$. This is not surprising, because No.~5 is very luminous with $M_{\rm UV} \approx -22$ mag and No.~13 is also luminous ($M_{\rm UV} \approx -20$ mag) with a very high \lya EW ($\sim190$ \AA). The two galaxies also have high stellar masses, relatively large ages, and relatively high dust extinction. No.~13 has been observed by JWST NIR-Cam and will be further discussed later. 

Observational studies of stellar populations in \lya galaxies have suggested that young and low-mass galaxies tend to hold \lya lines with large EWs \citep[e.g.,][]{Ono_2010,Hayes_2023,Roy_2023}. This can be explained by increasing gas and dust content in more massive galaxies and thus lower \lya escape fractions, or intrinsically low \lya production by more evolved populations. The middle panel of Figure~\ref{seds} shows that our galaxies generally follow this trend that less massive galaxies having higher \lya EWs

\subsection{Galaxy Morphology} \label{sec:image}

\begin{figure}[!t]
\centering
\includegraphics[width=0.5\textwidth]{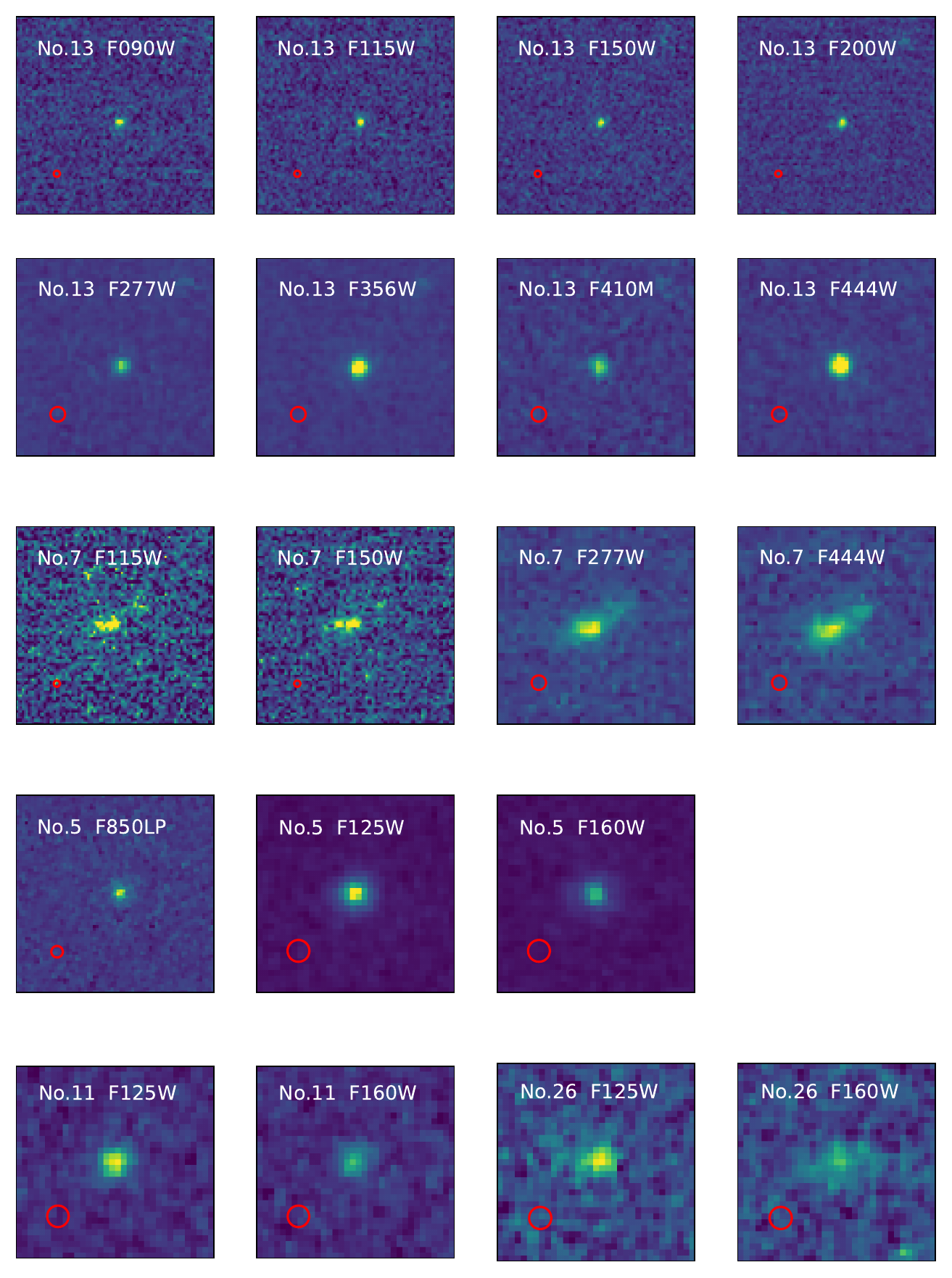}
\caption{ JWST NIRCam and HST ACS and WFC3 cutout images of five LBGs. The image size is $2'' \times 2''$. The red circles in each stamp image indicate the FWHM of the PSF. \label{jwst_hst_image}}
\end{figure}

Two LBGs in our sample (No.~7 and No.~13) have high-resolution JWST NIRCam images, and three LBGs (No.~5, No.~11, and No.~26) have HST WFC3 near-IR images. The JWST images are retrieved from the Mikulski Archive for Space Telescopes (MAST) and then reduced by the JWST Calibration Pipeline (version 1.11.4). We basically follow the procedure of the CEERS team for the image reduction \citep{Finkelstein_2023_CEERS}. The final resampled and combined images have a pixel scale of $0\farcs02$ for the short wavelengths (SW; F090W, F115W, F150W, and F200W) and $0\farcs04$ for the long wavelengths (LW; F277W, F356W, F410M, and F444W). The HST data of the three LBGs are from the CANDELS survey and the images are downloaded from the website\footnote{\url{https://archive.stsci.edu/hlsp/candels}}. The cutout images of the five LBGs are shown in Figure~\ref{jwst_hst_image}. 

We measure morphological parameters for the five LBGs using GALFIT. For either of the two LBGs with NIRCam images, we stack its SW images and its LW images to produce two combined images. This is to increase S/N, assuming that the object morphology and image quality are similar in individual input bands. Likewise, we stack the F125W and F160W images for each of the three LBGs with HST data. A PSF is obtained for each combined image by stacking several unsaturated stars in the same image, and the resultant FWHM of PSFs are shown as the red circles in Figure~\ref{jwst_hst_image}. The FWHM values of the PSFs in the NIRCam SW and LW bands are about $0\farcs063$ and $0\farcs148$ ($\approx0.36$ and 0.84 kpc at $z=6$) , respectively. The PSF FWHM of the HST images is about  $0\farcs223$ ($\approx1.3$ kpc at $z=6$).

We then use GALFIT to model the galaxies on the stacked images, and constrain a few basic parameters, including the effective radius $R_e$ along the major axis, $\rm{S\acute{e}rsic}$ index $n$ (allowed to vary between 0.3 and 4), and projected minor-to-major axis ratio ($b/a$). We find that all galaxies except No.~7 can be well fitted by one $\rm{S\acute{e}rsic}$ profile. No.~7 LBG seems to have two separate components and will be further discussed later. The fitting results of the $\rm{S\acute{e}rsic}$ profiles are reported in Table~\ref{galfit_fit}.

\begin{deluxetable*}{ccccc}
\tablewidth{\textwidth} 
\tablenum{3}
\tablecaption{Galaxy morphological parameters \label{galfit_fit}}
\tablehead{
\colhead{No.} & \colhead{bands}  & \colhead{$n$} & \colhead{$R_e$(kpc)} & \colhead{$b/a$} 
}
\startdata
\multirow{2}{*}{13} & F090W, F115W, F150W, F200W & $ 1.76 \pm 0.45 $&$ 0.25 \pm 0.03 $&$ 0.67 \pm 0.08 $ \\
                    & F277W, F356W, F410M, F444W & $ 1.10 \pm 0.27 $&$ 0.38 \pm 0.03 $&$ 0.99 \pm 0.12 $ \\
\multirow{2}{*}{$7^a$} & F115W, F150W & $ 0.49 \pm 0.13 $&$ 0.59 \pm 0.04 $&$ 0.51 \pm 0.04 $ \\
                       & F277W, F444W & $ 0.67 \pm 0.11 $&$ 1.02 \pm 0.05 $&$ 0.61 \pm 0.04 $ \\
\multirow{1}{*}{5}  & F125W, F160W &$ 1.18 \pm 0.19 $&$ 0.94 \pm 0.09 $&$ 0.92 \pm 0.08 $ \\
\multirow{1}{*}{11} & F125W, F160W &$ 1.00 \pm 0.29 $&$ 1.16 \pm 0.11 $&$ 0.59 \pm 0.09 $ \\
\multirow{1}{*}{26} & F125W, F160W &$ 0.44 \pm 0.19 $&$ 1.42 \pm 0.20 $&$ 0.70 \pm 0.11 $ \\
\enddata
\tablecomments{a. The parameters of $\rm{S\acute{e}rsic}$ profile of the main component is reported.}

\end{deluxetable*}

Figure~\ref{jwst_hst_image} and Table~\ref{galfit_fit} show that most galaxies in our sample are compact. For example, No.~13 is barely resolved. Its  effective radius is $R_e=0.25$~kpc in the SW bands and $R_e=0.38$~kpc in the LW bands, making it one of the most compact galaxies in this redshift and luminosity range (see, e.g., \citealt{sun_2023_structure}). Its large \lya EW, high SFR, and compact size is consistent with Green Pea (GP) galaxies found at low redshift \citep{Cardamone_2009_GP_1st} that likely have compact star-forming regions embedded in diffuse, older stellar components \citep{Amorin_2012_GP_pop,Clarke_2021_GP_pop}.
The GP galaxies are thought to have numerous counterparts at the reionization epoch and some of them may have been discovered recently by JWST \citep{Hall_2023_GP_JWST_Nat,Rhoads_2023_GP_JWST}. Studies of local GP galaxies have provided valuable information on how \lya photons escape from ISM and CGM, which help us better understand their high-redshift counterparts \citep[e.g.,][]{Henry_2015_GP,Yang_2017_GP}. Another possible explanation for the compactness of the No. 13 galaxy is that this galaxy holds an AGN that partly powers its strong \lya emission.

No.~7 LBG has a quite extended structure and appears to show two components, including a major component and a much weaker component to the north-west of the major one. The major component is more than five times brighter than the minor component in all four bands, so the major component dominates the radiation in these bands. 
It is unclear whether the weak component is part of the galaxy or a foreground object. In addition, mergers and  interacting systems occur more frequently at high redshift \citep{Rodriguez_2015}, so it is possible that No.~7 LBG is a merger. To verify that the two components belong to the same source, we estimate the probability that the distance between any two random sources is less than the distance between the two components of No.~7 LBG ($0.36''$). Based on the NIRCam images used above, we find that the probability is only 0.02, suggesting that the minor component is unlikely a foreground object. Therefore, we use an additional single $\rm{S\acute{e}rsic}$ profile to model the minor component of No.~7, and we obtain a much better fitting result. Table~\ref{galfit_fit} shows the result for the major component.

                                                                                                                                                                                                                                         
\section{Discussion} \label{sec:disc}

\subsection{Fraction of \lya emitters in LBGs} \label{sec:Lyafrac}

The fraction of LAEs among LBGs is an important tracer of the IGM state during the epoch of reionization, and many works have been done to constrain the fraction of the neutral hydrogen and its redshift evolution. This method is also called the \lya visibility test. Previous studies have shown that the fraction of LAEs in LBG samples increases steadily from low redshift to $z\sim6$, and drops dramatically towards higher redshift \citep[e.g.,][]{stark_keck_2010,stark_keck_2011,ono_spectroscopic_2012,Bian_2015_z7}. It is difficult to explain this using a sudden change in the physical properties of galaxies, but it can be easily explained by the increase of the fraction of the neutral hydrogen in the IGM that attenuates \lya emission. Before we calculate the LAE fraction in our LBG sample, we would like to clarify the definitions of LAEs and LBGs since the definitions are slightly different in the literature. LBGs mean galaxies selected by the Lyman-break technique, regardless of the existence of \lya emission. For LBGs at $z\ge6$, they are usually very faint, and spectroscopic identifications primarily rely on their \lya emission. This means that spectroscopically confirmed LBGs typically have strong \lya emission, and also means that the vast majority of the known LBGs at $z\ge6$ are not spectroscopically confirmed (or just candidates). LAEs are galaxies with strong \lya emission (typically EW $\ge25$ or 50 \AA). Narrowband-selected galaxies are mostly LAEs by definition. If an LBG has strong \lya emission, it is also an LAE, meaning that most spectroscopically confirmed $z\ge6$ LBGs to date are also LAEs. In the following analyses, we use EW $\ge25$ \AA\ to define LAEs.

We measure the LAE fraction among LBGs at $z \sim 6$ using our spectroscopically confirmed LBG sample. As mentioned earlier, our LBG selection criteria are slightly looser than those used in previous studies, which may result in a slightly higher contamination rate. Therefore, our result can be regarded as a lower limit. Because of the presence of strong OH skylines in the wavelength range that we probe, the identification of the \lya line is incomplete. 
We run simulations to estimate the completeness for the \lya line identification in the spectra. We generate a grid of model \lya lines with a range of redshifts $z=5.3$--$6.8$ and line strengths $L(Ly\alpha)=10^{42.5}$--$10^{43.5}\ \mathrm{erg/s}$ using the \lya template from \citet{ning_magellan_2020}. For each spectrum without a \lya detection, we insert an artificial \lya line. We then search for these lines using the same method that was used for real spectra in Section~\ref{subsec:identification}. The derived completeness function is the recovery rate of the lines in each redshift and \lya luminosity bin.

\begin{figure}[!t]
\centering
\includegraphics[width=0.45\textwidth]{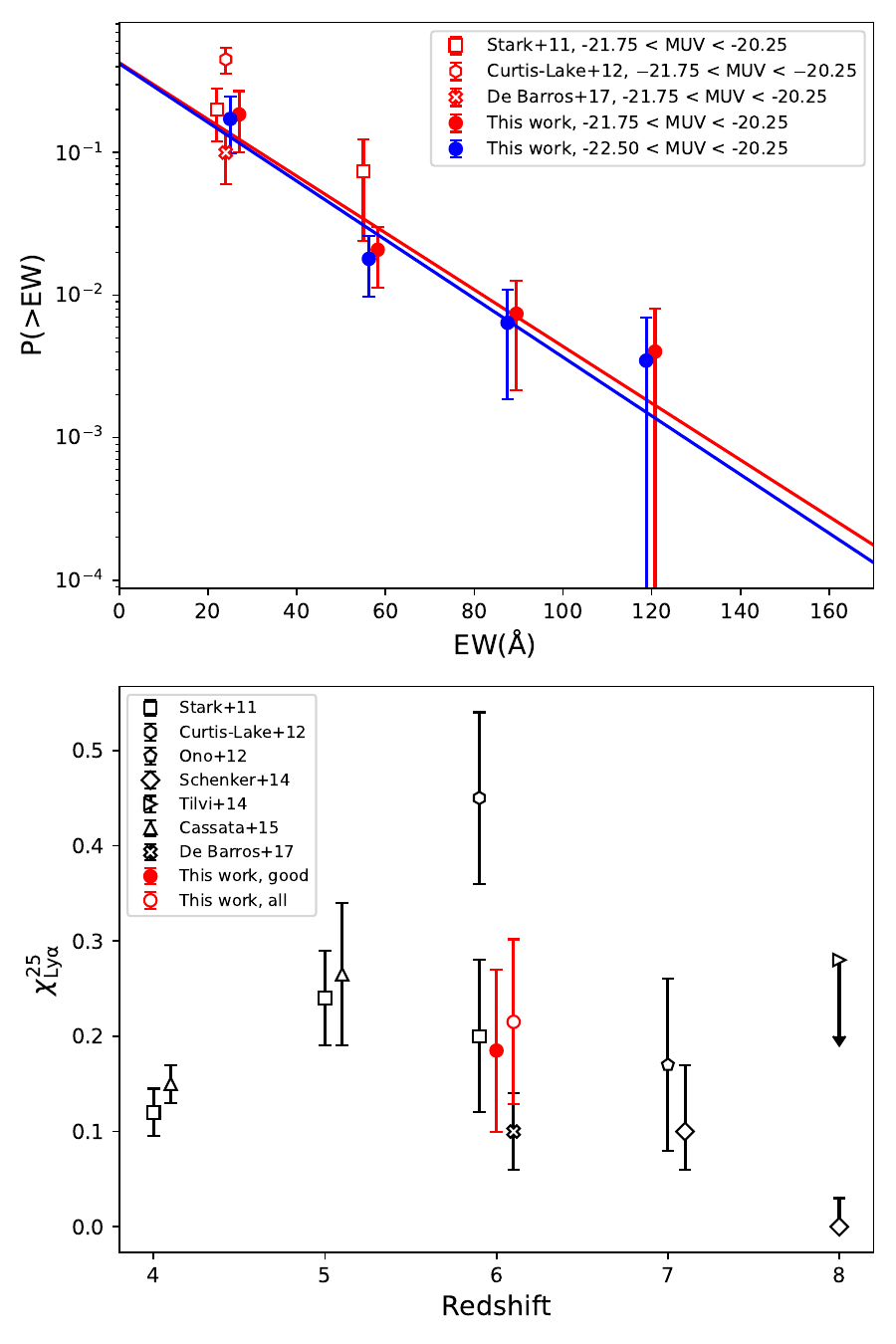}
\caption{Upper panel: cumulative distribution of the \lya rest-frame EWs P($>\mathrm{EW}$) for the $z \sim 6$ LBGs. The red symbols represent LBGs with $-21.75 < M_\mathrm{UV} < -20.25$ and the blue circles represent all LBGs within $-21.75 < M_\mathrm{UV} < -20.25$. The results are consistent with previous measurements.  The lines are the best fits to the distributions.
Lower panel: fraction of LAEs with $\mathrm{EW}>25$~\AA ($\chi^{25}_\mathrm{Ly\alpha}$) for $-21.75 < M_\mathrm{UV} < -20.25$ galaxies at $4 \leq z \leq 8$. The red filled circles are calculated from the `good' LBG sample and the red open circles are calculated from both `good' and `possible' samples. The results are compared with previous studies.  \label{LAE_frac}}
\end{figure}

We then apply the completeness correction to the LAE fraction and the result is shown in Figure~\ref{LAE_frac}. The upper panel of the figure shows the cumulative distribution of the \lya rest-frame EWs in the `good' LBG sample. The fraction in a fainter subsample ($-21.75 < M_\mathrm{UV} < 20.25$ mag) is marginally higher than that in the whole LBG sample ($-21.75 < M_\mathrm{UV} < -20.25$ mag). The results are generally consistent with previous studies \citep{pentericci_spectroscopic_2011,stark_keck_2011,curtis-lake_remarkably_2012,DeBarros_2017}. 
 
In the lower panel of Figure~\ref{LAE_frac}, we show the LAE fractions $\chi_\mathrm{Ly\alpha}^{25}$ in both `good' and the whole samples, and compare them with previous studies \citep{stark_keck_2011,curtis-lake_remarkably_2012,ono_spectroscopic_2012,Schenker_2014,tilvi_rapid_2014,cassata_vimos_2015,DeBarros_2017} in a redshift range of $4-8$.  The UV magnitude range of the LBGs is $-21.75 < M_\mathrm{UV} < -20.25$ mag, which corresponds to the fainter part of our sample and the brighter part of the previous samples. As seen from the figure, our result is broadly consistent with previous results, albeit with a large scatter. The large scatter and error bars reflect complex sample bias and incompleteness.
The figure also shows that $\chi_\mathrm{Ly\alpha}^{25}$ increases from lower redshifts to $z\sim6$ and then decreases towards higher redshifts, consistent with the previous claim. The mild decline of $\chi_\mathrm{Ly\alpha}^{25}$ may suggest that the evolution of the IGM opacity from $z\sim 7$ to $z\sim 6$ is not very dramatic, if the \lya visibility mostly depends on the IGM state. Nevertheless, larger and more complete samples are needed.

\subsection{SED modeling with strong \lya emission} \label{sec:SED_lyaNB}

In Section~\ref{sec:SED} when we modeled the SEDs of the LBGs, we did not consider the \lya emission due to its complexity. Here we discuss how the results change if we consider the \lya emission in our SED modeling. As has been pointed out, the SED modeling for $z \sim 6$ galaxies has a strong degeneracy between young galaxies with prominent nebular emission and old galaxies with strong Balmer breaks \citep[e.g.,][]{jiang_physical_2015}. To include the \lya emission, we make a pseudo narrowband filter with a spectral resolving power of $R=100$ at the wavelength of \lya for each LBG. The narrowband photometry is then calculated from the \lya flux and implemented for SED modeling. In the modeling procedure, the ionization parameter $U$, gas metallicity in the nebular emission model, and the ratio of continuum to line reddening are considered as free parameters. 

\begin{figure}[!t]
\centering
\includegraphics[width=0.4\textwidth]{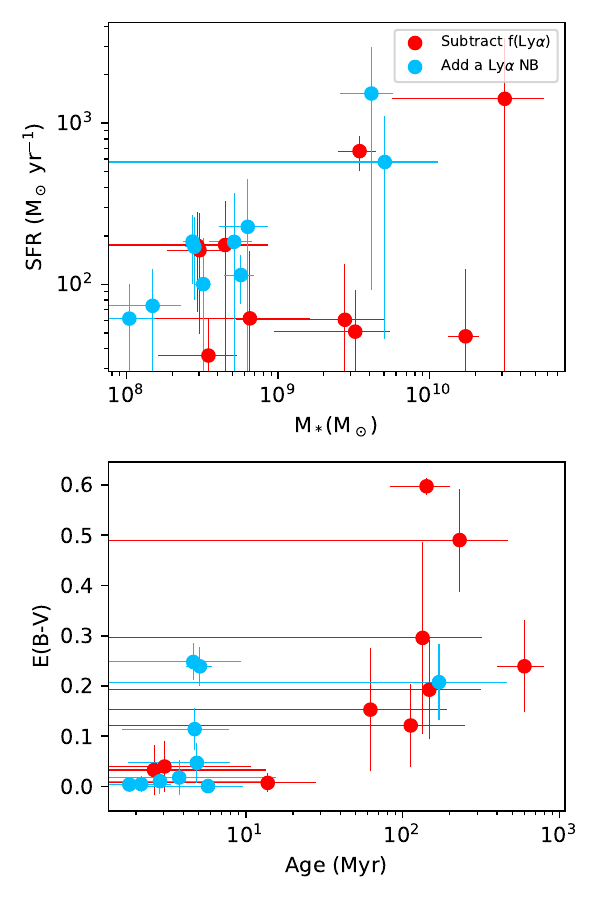}
\caption{ Comparison of the stellar populations obtained by two SED modeling methods, with and without \lya constraints (see details in Section 5.2).  \label{properties_compare}}
\end{figure}

The derived properties of the stellar populations and the comparison with the previous results from Section~\ref{sec:SED} are shown in Figure~\ref{properties_compare}. The best-fit $\chi_r^2$ values of the two methods are about the same.
Figure~\ref{properties_compare} clearly shows that when the \lya emission is considered in the SED modeling procedure, these galaxies are much younger with significantly lower stellar masses. The reason is straightforward. In order to produce strong \lya emission seen in the galaxies, the stellar populations must be very young with high SFRs, which further results in lower stellar masses. It should be noted, however, that it is difficult to decide how to include the \lya emission for SED modeling due to the complexity of the \lya emission. For example, the measurement of the \lya flux is almost always a lower limit, this is mainly because the resonant scattering of the \lya photons often forms a large diffuse halo, but the actual measurement is usually done on the central part. It is also because the \lya escape fraction varies a lot among galaxies. In addition, there are other sources of uncertainties that are particularly important for $z\ge6$ galaxies, including the variation of the IGM absorption along the line-of-sights.

\subsection{Search for AGN activity} \label{sec:noAGN}

The LBGs in our sample have strong \lya and UV continuum emission. In addition, some of them are morphologically very compact. It is possible that they harbor a central AGN that partly contributes to the total emission. We search for possible evidence of AGN activity in these galaxies. First of all, they do not have typical Type I AGN components, as seen from their narrow \lya emission lines. Figure~\ref{FWHM} shows that the stacked \lya line displays a narrow profile with a velocity dispersion of $\Delta v \sim 250$~km/s (IGM absorption is neglected here), which is much smaller than those in typical broad-line AGN ($\ge2000$ km/s).

The wavelength coverage of our spectra is from 7600~\AA\ to 9600~\AA, corresponding to a rest-frame range of 1090-1370~\AA\ for $z\sim6$ objects. This range covers the N\,{\scriptsize V} 1240~\AA\ emission line that is often regarded as evidence of the presence of AGN. We do not detect any N\,{\scriptsize V} emission in the combined spectrum of the LBGs. For a typical AGN, however, the N\,{\scriptsize V} flux is only a few percent of the \lya flux \citep[e.g.,][]{Berk_2001_sdss_composite_spec}, so even if any of our LBGs contain an AGN component, its N\,{\scriptsize V} flux is below our detection limit. 

We also check these galaxies in deep Chandra X-ray images, including the Chandra Deep Field-South field (CDF-S) \citep{Luo_2017_cdfs}, the UKIDSS Ultra Deep Survey Field \citep{Kocevski_2018_XUDS}, and the Chandra Cosmos Legacy Survey field \citep{Civano_2016_chandracosmos}. None of them is individually detected. X-ray detections were actually excluded in the first steps of the target selection. 
The No.5 and No.6 LBGs are in the 7~Ms CDF-S field, and their stacked image have an effective exposure time of over 12~Ms. For the stacked 0.5--7~keV image, we measure the X-ray photon counts in a $1\farcs5$ radius aperture, which encloses $90\%$ of the total energy at 6.4~keV. The background counts are computed in an annular region with the inner and outer radii of $4\arcsec$ and $8\arcsec$. The $1\sigma$ error of the net counts is calculated based on the Poisson errors on the extracted source and background counts. The calculated $1\sigma$ upper limit is 6.3 counts. To estimate the upper limit of a possible X-ray detection, we assume a typical type 1 quasar with a similar UV magnitude  at the same redshift as the No.5 and No.6 LBGs ($z\sim5.7$, $M_{\rm UV}=-22$~ mag). We also assume the  \citet{Shen_2020_sed} SED template for the quasar. We adopt the Galactic neutral hydrogen column density along the line-of-sight to the CDF-S $n_\mathrm{H}=8.8 \times 10^{19} \mathrm{cm}^{-2}$ \citep{Stark_1992_NH}. No other absorption is applied. In this case, the 12~Ms exposure would yield a $2\sigma$ detection (19.2 counts in 0.5--7~keV band) for this quasar. This simple calculation means that AGN contributes less than one-third ($1\sigma$ upper limit) of the total luminosity in the two LBGs. Therefore, we are not able to rule out the possibility of weaker AGN activities in these galaxies. The stacked images in another two fields are much shallower and thus cannot provide useful constraints.


\section{Summary} \label{sec:summary}

We have presented a sample of 45 spectroscopically confirmed LBGs at $z\sim 6$ in four well-studied fields, including SXDS, A370, ECDFS, and COSMOS. This sample is one of the largest samples of spectroscopically confirmed LBGs at this redshift. The LBG candidates were selected as $i$-band dropout objects from deep broadband images. The spectroscopic observations were carried out using M2FS on the Magellan Clay telescope. We identified the \lya lines of the galaxies based on their 1D and 2D M2FS spectra. Their redshifts and \lya luminosities were measured by fitting a composite \lya line template to the individual 1D spectra. The absolute UV magnitude and UV slope were calculated by fitting a power-law to the available multi-wavelength photometric data. The UV absolute magnitudes of these LBGs are between --22.5 and --19.0 mag, corresponding to $0.3-3.4$ times the characteristic luminosity of galaxies at $z\sim 6$, with a \lya EW range from $\sim$10~\AA\ to $\sim$400~\AA. They represent the most UV luminous galaxies known at $z\sim 6$.

The SED modeling of ten LBGs in the sample reveals that they have a wide range of stellar masses and ages with high SFRs. We note that whether to consider the \lya emission in the SED modeling procedure has a significant impact on the measurement of the stellar populations in these galaxies, indicating the complexity of the \lya emission. Five LBGs have high-resolution JWST NIRCam images or HST WFC3 near-IR images. Four of them show compact morphology, and one appears to have two components that might be a merger. 
We calculated the LAE fraction among the photometrically selected LBGs after correcting for the Ly$\alpha$-identification completeness. The fraction of $\sim 0.2$ is consistent with previous works, and it supports a moderate evolution of the IGM opacity at the end of cosmic reionization when previous results at lower and higher redshifts were combined. Finally, we did not find evidence of strong AGN activities in these galaxies.


\begin{acknowledgments}

We acknowledge support from the National Science Foundation of China (12225301), the National Key R\&D Program of China (2022YFF0503401), and the China Manned Space Project with No. CMS-CSST-2021-A05 and CMS-CSST-2021-A07.

\end{acknowledgments}

\facility{Magellan: Clay (M2FS).}
\software{Astropy \citep{astropy:2013, astropy:2018, astropy:2022}, CIGALE \citep{cigale:2009}, GALFIT \citep{galfit:2002,galfit:2010}}

\bibliography{ms}

\begin{thebibliography}{}
\expandafter\ifx\csname natexlab\endcsname\relax\def\natexlab#1{#1}\fi
\providecommand{\url}[1]{\href{#1}{#1}}
\providecommand{\dodoi}[1]{doi:~\href{http://doi.org/#1}{\nolinkurl{#1}}}
\providecommand{\doeprint}[1]{\href{http://ascl.net/#1}{\nolinkurl{http://ascl.net/#1}}}
\providecommand{\doarXiv}[1]{\href{https://arxiv.org/abs/#1}{\nolinkurl{https://arxiv.org/abs/#1}}}

\bibitem[{Aihara {et~al.}(2022)Aihara, AlSayyad, Ando, Armstrong, Bosch, Egami,
  Furusawa, Furusawa, Harasawa, Harikane, Hsieh, Ikeda, Ito, Iwata, Kodama,
  Koike, Kokubo, Komiyama, Li, Liang, Lin, Lupton, Lust, MacArthur, Mawatari,
  Mineo, Miyatake, Miyazaki, More, Morishima, Murayama, Nakajima, Nakata,
  Nishizawa, Oguri, Okabe, Okura, Ono, Osato, Ouchi, Pan, Malag{\'{o}}n, Price,
  Reed, Rykoff, Shibuya, Simunovic, Strauss, Sugimori, Suto, Suzuki, Takada,
  Takagi, Takata, Takita, Tanaka, Tang, Taranu, Terai, Toba, Turner, Uchiyama,
  Vijarnwannaluk, Waters, Yamada, Yamamoto, \& Yamashita}]{Aihara_2022_HSCSSP}
Aihara, H., AlSayyad, Y., Ando, M., {et~al.} 2022, Publications of the
  Astronomical Society of Japan, 74, 247, \dodoi{10.1093/pasj/psab122}

\bibitem[{{Amor{\'\i}n} {et~al.}(2012){Amor{\'\i}n}, {P{\'e}rez-Montero},
  {V{\'\i}lchez}, \& {Papaderos}}]{Amorin_2012_GP_pop}
{Amor{\'\i}n}, R., {P{\'e}rez-Montero}, E., {V{\'\i}lchez}, J.~M., \&
  {Papaderos}, P. 2012, \apj, 749, 185, \dodoi{10.1088/0004-637X/749/2/185}

\bibitem[{{Arrabal Haro} {et~al.}(2023){Arrabal Haro}, {Dickinson},
  {Finkelstein}, {Kartaltepe}, {Donnan}, {Burgarella}, {Carnall}, {Cullen},
  {Dunlop}, {Fern{\'a}ndez}, {Fujimoto}, {Jung}, {Krips}, {Larson}, {Papovich},
  {P{\'e}rez-Gonz{\'a}lez}, {Amor{\'\i}n}, {Bagley}, {Buat}, {Casey},
  {Chworowsky}, {Cohen}, {Ferguson}, {Giavalisco}, {Huertas-Company},
  {Hutchison}, {Kocevski}, {Koekemoer}, {Lucas}, {McLeod}, {McLure}, {Pirzkal},
  {Trump}, {Weiner}, {Wilkins}, \& {Zavala}}]{Arrabal_Haro_2023_2z10GLX}
{Arrabal Haro}, P., {Dickinson}, M., {Finkelstein}, S.~L., {et~al.} 2023, arXiv
  e-prints, arXiv:2303.15431, \dodoi{10.48550/arXiv.2303.15431}

\bibitem[{{Astropy Collaboration} {et~al.}(2013){Astropy Collaboration},
  {Robitaille}, {Tollerud}, {Greenfield}, {Droettboom}, {Bray}, {Aldcroft},
  {Davis}, {Ginsburg}, {Price-Whelan}, {Kerzendorf}, {Conley}, {Crighton},
  {Barbary}, {Muna}, {Ferguson}, {Grollier}, {Parikh}, {Nair}, {Unther},
  {Deil}, {Woillez}, {Conseil}, {Kramer}, {Turner}, {Singer}, {Fox}, {Weaver},
  {Zabalza}, {Edwards}, {Azalee Bostroem}, {Burke}, {Casey}, {Crawford},
  {Dencheva}, {Ely}, {Jenness}, {Labrie}, {Lim}, {Pierfederici}, {Pontzen},
  {Ptak}, {Refsdal}, {Servillat}, \& {Streicher}}]{astropy:2013}
{Astropy Collaboration}, {Robitaille}, T.~P., {Tollerud}, E.~J., {et~al.} 2013,
  \aap, 558, A33, \dodoi{10.1051/0004-6361/201322068}

\bibitem[{{Astropy Collaboration} {et~al.}(2018){Astropy Collaboration},
  {Price-Whelan}, {Sip{\H{o}}cz}, {G{\"u}nther}, {Lim}, {Crawford}, {Conseil},
  {Shupe}, {Craig}, {Dencheva}, {Ginsburg}, {Vand erPlas}, {Bradley},
  {P{\'e}rez-Su{\'a}rez}, {de Val-Borro}, {Aldcroft}, {Cruz}, {Robitaille},
  {Tollerud}, {Ardelean}, {Babej}, {Bach}, {Bachetti}, {Bakanov}, {Bamford},
  {Barentsen}, {Barmby}, {Baumbach}, {Berry}, {Biscani}, {Boquien}, {Bostroem},
  {Bouma}, {Brammer}, {Bray}, {Breytenbach}, {Buddelmeijer}, {Burke},
  {Calderone}, {Cano Rodr{\'\i}guez}, {Cara}, {Cardoso}, {Cheedella}, {Copin},
  {Corrales}, {Crichton}, {D'Avella}, {Deil}, {Depagne}, {Dietrich}, {Donath},
  {Droettboom}, {Earl}, {Erben}, {Fabbro}, {Ferreira}, {Finethy}, {Fox},
  {Garrison}, {Gibbons}, {Goldstein}, {Gommers}, {Greco}, {Greenfield},
  {Groener}, {Grollier}, {Hagen}, {Hirst}, {Homeier}, {Horton}, {Hosseinzadeh},
  {Hu}, {Hunkeler}, {Ivezi{\'c}}, {Jain}, {Jenness}, {Kanarek}, {Kendrew},
  {Kern}, {Kerzendorf}, {Khvalko}, {King}, {Kirkby}, {Kulkarni}, {Kumar},
  {Lee}, {Lenz}, {Littlefair}, {Ma}, {Macleod}, {Mastropietro}, {McCully},
  {Montagnac}, {Morris}, {Mueller}, {Mumford}, {Muna}, {Murphy}, {Nelson},
  {Nguyen}, {Ninan}, {N{\"o}the}, {Ogaz}, {Oh}, {Parejko}, {Parley}, {Pascual},
  {Patil}, {Patil}, {Plunkett}, {Prochaska}, {Rastogi}, {Reddy Janga},
  {Sabater}, {Sakurikar}, {Seifert}, {Sherbert}, {Sherwood-Taylor}, {Shih},
  {Sick}, {Silbiger}, {Singanamalla}, {Singer}, {Sladen}, {Sooley},
  {Sornarajah}, {Streicher}, {Teuben}, {Thomas}, {Tremblay}, {Turner},
  {Terr{\'o}n}, {van Kerkwijk}, {de la Vega}, {Watkins}, {Weaver}, {Whitmore},
  {Woillez}, {Zabalza}, \& {Astropy Contributors}}]{astropy:2018}
{Astropy Collaboration}, {Price-Whelan}, A.~M., {Sip{\H{o}}cz}, B.~M., {et~al.}
  2018, \aj, 156, 123, \dodoi{10.3847/1538-3881/aabc4f}

\bibitem[{{Astropy Collaboration} {et~al.}(2022){Astropy Collaboration},
  {Price-Whelan}, {Lim}, {Earl}, {Starkman}, {Bradley}, {Shupe}, {Patil},
  {Corrales}, {Brasseur}, {N{"o}the}, {Donath}, {Tollerud}, {Morris},
  {Ginsburg}, {Vaher}, {Weaver}, {Tocknell}, {Jamieson}, {van Kerkwijk},
  {Robitaille}, {Merry}, {Bachetti}, {G{"u}nther}, {Aldcroft},
  {Alvarado-Montes}, {Archibald}, {B{'o}di}, {Bapat}, {Barentsen}, {Baz{'a}n},
  {Biswas}, {Boquien}, {Burke}, {Cara}, {Cara}, {Conroy}, {Conseil}, {Craig},
  {Cross}, {Cruz}, {D'Eugenio}, {Dencheva}, {Devillepoix}, {Dietrich},
  {Eigenbrot}, {Erben}, {Ferreira}, {Foreman-Mackey}, {Fox}, {Freij}, {Garg},
  {Geda}, {Glattly}, {Gondhalekar}, {Gordon}, {Grant}, {Greenfield}, {Groener},
  {Guest}, {Gurovich}, {Handberg}, {Hart}, {Hatfield-Dodds}, {Homeier},
  {Hosseinzadeh}, {Jenness}, {Jones}, {Joseph}, {Kalmbach}, {Karamehmetoglu},
  {Ka{l}uszy{'n}ski}, {Kelley}, {Kern}, {Kerzendorf}, {Koch}, {Kulumani},
  {Lee}, {Ly}, {Ma}, {MacBride}, {Maljaars}, {Muna}, {Murphy}, {Norman},
  {O'Steen}, {Oman}, {Pacifici}, {Pascual}, {Pascual-Granado}, {Patil},
  {Perren}, {Pickering}, {Rastogi}, {Roulston}, {Ryan}, {Rykoff}, {Sabater},
  {Sakurikar}, {Salgado}, {Sanghi}, {Saunders}, {Savchenko}, {Schwardt},
  {Seifert-Eckert}, {Shih}, {Jain}, {Shukla}, {Sick}, {Simpson},
  {Singanamalla}, {Singer}, {Singhal}, {Sinha}, {Sip{H{o}}cz}, {Spitler},
  {Stansby}, {Streicher}, {{{S}}umak}, {Swinbank}, {Taranu}, {Tewary},
  {Tremblay}, {Val-Borro}, {Van Kooten}, {Vasovi{'c}}, {Verma}, {de Miranda
  Cardoso}, {Williams}, {Wilson}, {Winkel}, {Wood-Vasey}, {Xue}, {Yoachim},
  {Zhang}, {Zonca}, \& {Astropy Project Contributors}}]{astropy:2022}
{Astropy Collaboration}, {Price-Whelan}, A.~M., {Lim}, P.~L., {et~al.} 2022,
  apj, 935, 167, \dodoi{10.3847/1538-4357/ac7c74}

\bibitem[{{Atek} {et~al.}(2023){Atek}, {Chemerynska}, {Wang}, {Furtak},
  {Weibel}, {Oesch}, {Weaver}, {Labb{\'e}}, {Bezanson}, {van Dokkum}, {Zitrin},
  {Dayal}, {Williams}, {Nannayakkara}, {Price}, {Brammer}, {Goulding}, {Leja},
  {Marchesini}, {Nelson}, {Pan}, \& {Whitaker}}]{Atek_2023_jwst_dropout}
{Atek}, H., {Chemerynska}, I., {Wang}, B., {et~al.} 2023, \mnras, 524, 5486,
  \dodoi{10.1093/mnras/stad1998}

\bibitem[{{Bian} {et~al.}(2015){Bian}, {Stark}, {Fan}, {Jiang}, {Cl{\'e}ment},
  {Egami}, {Frye}, {Green}, {McGreer}, \& {Cai}}]{Bian_2015_z7}
{Bian}, F., {Stark}, D.~P., {Fan}, X., {et~al.} 2015, \apj, 806, 108,
  \dodoi{10.1088/0004-637X/806/1/108}

\bibitem[{Bouwens {et~al.}(2014)Bouwens, Illingworth, Oesch, Labb{\'e}, van
  Dokkum, Trenti, Franx, Smit, Gonzalez, \& Magee}]{bouwens_uv-continuum_2014}
Bouwens, R.~J., Illingworth, G.~D., Oesch, P.~A., {et~al.} 2014, The
  Astrophysical Journal, 793, 115, \dodoi{10.1088/0004-637X/793/2/115}

\bibitem[{Bouwens {et~al.}(2021)Bouwens, Oesch, Stefanon, Illingworth,
  Labb{\'e}, Reddy, Atek, Montes, Naidu, Nanayakkara, Nelson, \&
  Wilkins}]{bouwens_new_2021}
Bouwens, R.~J., Oesch, P.~A., Stefanon, M., {et~al.} 2021, The Astronomical
  Journal, 162, 47, \dodoi{10.3847/1538-3881/abf83e}

\bibitem[{{Bruzual} \& {Charlot}(2003)}]{Bruzual_2003_bc03}
{Bruzual}, G., \& {Charlot}, S. 2003, \mnras, 344, 1000,
  \dodoi{10.1046/j.1365-8711.2003.06897.x}

\bibitem[{{Bunker} {et~al.}(2023){Bunker}, {Cameron}, {Curtis-Lake},
  {Jakobsen}, {Carniani}, {Curti}, {Witstok}, {Maiolino}, {D'Eugenio},
  {Looser}, {Willott}, {Bonaventura}, {Hainline}, {Uebler}, {Willmer},
  {Saxena}, {Smit}, {Alberts}, {Arribas}, {Baker}, {Baum}, {Bhatawdekar},
  {Bowler}, {Boyett}, {Charlot}, {Chen}, {Chevallard}, {Circosta}, {DeCoursey},
  {de Graaff}, {Egami}, {Eisenstein}, {Endsley}, {Ferruit}, {Giardino},
  {Hausen}, {Helton}, {Hviding}, {Ji}, {Johnson}, {Jones}, {Kumari}, {Laseter},
  {Luetzgendorf}, {Maseda}, {Nelson}, {Parlanti}, {Perna}, {Rawle}, {Rix},
  {Rieke}, {Robertson}, {Rodriguez Del Pino}, {Sandles}, {Scholtz}, {Sharpe},
  {Skarbinski}, {Stark}, {Sun}, {Tacchella}, {Topping}, {Villanueva},
  {Wallace}, {Williams}, \& {Woodrum}}]{Bunker_2023_jades}
{Bunker}, A.~J., {Cameron}, A.~J., {Curtis-Lake}, E., {et~al.} 2023, arXiv
  e-prints, arXiv:2306.02467, \dodoi{10.48550/arXiv.2306.02467}

\bibitem[{{Calzetti} {et~al.}(2000){Calzetti}, {Armus}, {Bohlin}, {Kinney},
  {Koornneef}, \& {Storchi-Bergmann}}]{Calzetti_2000_dust}
{Calzetti}, D., {Armus}, L., {Bohlin}, R.~C., {et~al.} 2000, \apj, 533, 682,
  \dodoi{10.1086/308692}

\bibitem[{{Cardamone} {et~al.}(2009){Cardamone}, {Schawinski}, {Sarzi},
  {Bamford}, {Bennert}, {Urry}, {Lintott}, {Keel}, {Parejko}, {Nichol},
  {Thomas}, {Andreescu}, {Murray}, {Raddick}, {Slosar}, {Szalay}, \&
  {Vandenberg}}]{Cardamone_2009_GP_1st}
{Cardamone}, C., {Schawinski}, K., {Sarzi}, M., {et~al.} 2009, \mnras, 399,
  1191, \dodoi{10.1111/j.1365-2966.2009.15383.x}

\bibitem[{{Casey} {et~al.}(2022){Casey}, {Kartaltepe}, {Drakos}, {Franco},
  {Harish}, {Paquereau}, {Ilbert}, {Rose}, {Cox}, {Nightingale}, {Robertson},
  {Silverman}, {Koekemoer}, {Massey}, {McCracken}, {Rhodes}, {Akins},
  {Amvrosiadis}, {Arango-Toro}, {Bagley}, {Bongiorno}, {Capak}, {Champagne},
  {Chartab}, {Chavez Ortiz}, {Chworowsky}, {Cooke}, {Cooper}, {Darvish},
  {Ding}, {Faisst}, {Finkelstein}, {Fujimoto}, {Gentile}, {Gillman}, {Gould},
  {Gozaliasl}, {Hayward}, {He}, {Hemmati}, {Hirschmann}, {Jahnke}, {Jin},
  {Khostovan}, {Kokorev}, {Lambrides}, {Laigle}, {Larson}, {Leung}, {Liu},
  {Liaudat}, {Long}, {Magdis}, {Mahler}, {Mainieri}, {Manning}, {Maraston},
  {Martin}, {McCleary}, {McKinney}, {McPartland}, {Mobasher}, {Pattnaik},
  {Renzini}, {Rich}, {Sanders}, {Sattari}, {Scognamiglio}, {Scoville}, {Sheth},
  {Shuntov}, {Sparre}, {Suzuki}, {Talia}, {Toft}, {Trakhtenbrot}, {Urry},
  {Valentino}, {Vanderhoof}, {Vardoulaki}, {Weaver}, {Whitaker}, {Wilkins},
  {Yang}, \& {Zavala}}]{Caitlin_2022_cosmosweb}
{Casey}, C.~M., {Kartaltepe}, J.~S., {Drakos}, N.~E., {et~al.} 2022, arXiv
  e-prints, arXiv:2211.07865, \dodoi{10.48550/arXiv.2211.07865}

\bibitem[{Cassata {et~al.}(2015)Cassata, Tasca, Le~F{\`e}vre, Lemaux, Garilli,
  Le~Brun, Maccagni, Pentericci, Thomas, Vanzella, Zamorani, Zucca, Amorin,
  Bardelli, Capak, Cassar{\`a}, Castellano, Cimatti, Cuby, Cucciati, de~la
  Torre, Durkalec, Fontana, Giavalisco, Grazian, Hathi, Ilbert, Moreau,
  Paltani, Ribeiro, Salvato, Schaerer, Scodeggio, Sommariva, Talia, Taniguchi,
  Tresse, Vergani, Wang, Charlot, Contini, Fotopoulou, Koekemoer,
  L{\'o}pez-Sanjuan, Mellier, \& Scoville}]{cassata_vimos_2015}
Cassata, P., Tasca, L. A.~M., Le~F{\`e}vre, O., {et~al.} 2015, Astronomy \&
  Astrophysics, 573, A24, \dodoi{10.1051/0004-6361/201423824}

\bibitem[{Castellano {et~al.}(2017)Castellano, Pentericci, Fontana, Vanzella,
  Merlin, Barros, Amorin, Caputi, Cristiani, Finkelstein, Giallongo, Grazian,
  Koekemoer, Maiolino, Paris, Pilo, Santini, \& Yan}]{castellano_optical_2017}
Castellano, M., Pentericci, L., Fontana, A., {et~al.} 2017, The Astrophysical
  Journal, 839, 73, \dodoi{10.3847/1538-4357/aa696e}

\bibitem[{{Champagne} {et~al.}(2023){Champagne}, {Casey}, {Finkelstein},
  {Bagley}, {Cooper}, {Larson}, {Long}, \&
  {Wang}}]{Champagne_2023_jwst_dropout}
{Champagne}, J.~B., {Casey}, C.~M., {Finkelstein}, S.~L., {et~al.} 2023, \apj,
  952, 99, \dodoi{10.3847/1538-4357/acda8d}

\bibitem[{{Charlot} \& {Fall}(1993)}]{Stephane_1993_lya}
{Charlot}, S., \& {Fall}, S.~M. 1993, \apj, 415, 580, \dodoi{10.1086/173187}

\bibitem[{{Civano} {et~al.}(2016){Civano}, {Marchesi}, {Comastri}, {Urry},
  {Elvis}, {Cappelluti}, {Puccetti}, {Brusa}, {Zamorani}, {Hasinger},
  {Aldcroft}, {Alexander}, {Allevato}, {Brunner}, {Capak}, {Finoguenov},
  {Fiore}, {Fruscione}, {Gilli}, {Glotfelty}, {Griffiths}, {Hao}, {Harrison},
  {Jahnke}, {Kartaltepe}, {Karim}, {LaMassa}, {Lanzuisi}, {Miyaji}, {Ranalli},
  {Salvato}, {Sargent}, {Scoville}, {Schawinski}, {Schinnerer}, {Silverman},
  {Smolcic}, {Stern}, {Toft}, {Trakhtenbrot}, {Treister}, \&
  {Vignali}}]{Civano_2016_chandracosmos}
{Civano}, F., {Marchesi}, S., {Comastri}, A., {et~al.} 2016, \apj, 819, 62,
  \dodoi{10.3847/0004-637X/819/1/62}

\bibitem[{{Clarke} {et~al.}(2021){Clarke}, {Scarlata}, {Mehta}, {Keel},
  {Cardamone}, {Hayes}, {Adams}, {Dickinson}, {Fortson}, {Kruk}, {Lintott}, \&
  {Simmons}}]{Clarke_2021_GP_pop}
{Clarke}, L., {Scarlata}, C., {Mehta}, V., {et~al.} 2021, \apjl, 912, L22,
  \dodoi{10.3847/2041-8213/abf7cc}

\bibitem[{Curtis-Lake {et~al.}(2012)Curtis-Lake, McLure, Pearce, Dunlop,
  Cirasuolo, Stark, Almaini, Bradshaw, Chuter, Foucaud, \&
  Hartley}]{curtis-lake_remarkably_2012}
Curtis-Lake, E., McLure, R.~J., Pearce, H.~J., {et~al.} 2012, Monthly Notices
  of the Royal Astronomical Society, 422, 1425,
  \dodoi{10.1111/j.1365-2966.2012.20720.x}

\bibitem[{Curtis-Lake {et~al.}(2016)Curtis-Lake, McLure, Dunlop, Rogers,
  Targett, Dekel, Ellis, Faber, Ferguson, Grogin, Kocevski, Koekemoer, Lai,
  M{\'a}rmol-Queralt{\'o}, \& Robertson}]{curtis-lake_non-parametric_2016}
Curtis-Lake, E., McLure, R.~J., Dunlop, J.~S., {et~al.} 2016, Monthly Notices
  of the Royal Astronomical Society, 457, 440, \dodoi{10.1093/mnras/stv3017}

\bibitem[{{Curtis-Lake} {et~al.}(2023){Curtis-Lake}, {Carniani}, {Cameron},
  {Charlot}, {Jakobsen}, {Maiolino}, {Bunker}, {Witstok}, {Smit}, {Chevallard},
  {Willott}, {Ferruit}, {Arribas}, {Bonaventura}, {Curti}, {D'Eugenio},
  {Franx}, {Giardino}, {Looser}, {L{\"u}tzgendorf}, {Maseda}, {Rawle}, {Rix},
  {Rodr{\'\i}guez del Pino}, {{\"U}bler}, {Sirianni}, {Dressler}, {Egami},
  {Eisenstein}, {Endsley}, {Hainline}, {Hausen}, {Johnson}, {Rieke},
  {Robertson}, {Shivaei}, {Stark}, {Tacchella}, {Williams}, {Willmer},
  {Bhatawdekar}, {Bowler}, {Boyett}, {Chen}, {de Graaff}, {Helton}, {Hviding},
  {Jones}, {Kumari}, {Lyu}, {Nelson}, {Perna}, {Sandles}, {Saxena}, {Suess},
  {Sun}, {Topping}, {Wallace}, \& {Whitler}}]{Curtis_Lake_2023_4z10_13GLX}
{Curtis-Lake}, E., {Carniani}, S., {Cameron}, A., {et~al.} 2023, Nature
  Astronomy, 7, 622, \dodoi{10.1038/s41550-023-01918-w}

\bibitem[{{De Barros} {et~al.}(2017){De Barros}, {Pentericci}, {Vanzella},
  {Castellano}, {Fontana}, {Grazian}, {Conselice}, {Yan}, {Koekemoer},
  {Cristiani}, {Dickinson}, {Finkelstein}, \& {Maiolino}}]{DeBarros_2017}
{De Barros}, S., {Pentericci}, L., {Vanzella}, E., {et~al.} 2017, \aap, 608,
  A123, \dodoi{10.1051/0004-6361/201731476}

\bibitem[{Dunlop {et~al.}(2012)Dunlop, McLure, Robertson, Ellis, Stark,
  Cirasuolo, \& de~Ravel}]{dunlop_critical_2012}
Dunlop, J.~S., McLure, R.~J., Robertson, B.~E., {et~al.} 2012, Monthly Notices
  of the Royal Astronomical Society, 420, 901,
  \dodoi{10.1111/j.1365-2966.2011.20102.x}

\bibitem[{{Dunlop} {et~al.}(2021){Dunlop}, {Abraham}, {Ashby}, {Bagley},
  {Best}, {Bongiorno}, {Bouwens}, {Bowler}, {Brammer}, {Bremer}, {Calabro'},
  {Carnall}, {Castellano}, {Cirasuolo}, {Conselice}, {Cullen}, {Dave}, {Dayal},
  {Dekel}, {Dickinson}, {Duncan}, {Elbaz}, {Ellis}, {Ferguson}, {Ferrara},
  {Finkelstein}, {Fontana}, {Furlanetto}, {Fynbo}, {Gallerani}, {Gardner},
  {Giavalisco}, {Grazian}, {Grogin}, {Harikane}, {Hopkins}, {Ilbert},
  {Illingworth}, {Juneau}, {Jung}, {Kartaltepe}, {Kassin}, {Kauffmann},
  {Khochfar}, {Kirkpatrick}, {Kocevski}, {Koekemoer}, {Labbe}, {Laporte},
  {Larson}, {Lucas}, {Magee}, {Mason}, {McCracken}, {McLeod}, {McLure},
  {Merlin}, {Mesinger}, {Milvang-Jensen}, {Newman}, {Oesch}, {Ouchi},
  {Pacifici}, {Papovich}, {Peacock}, {Peeples}, {Pentericci}, {Perez-Gonzalez},
  {Pirzkal}, {Pope}, {Pye}, {Reddy}, {Robertson}, {Salvato}, {Santini},
  {Schaerer}, {Shapley}, {Simons}, {Smit}, {Smith}, {Snyder}, {Somerville},
  {Stanway}, {Stefanon}, {Tasca}, {Tikkanen}, {Tresse}, {Trump}, {Whitaker},
  {Wilkins}, {Wright}, {Wyithe}, {van Dokkum}, \& {van der
  Werf}}]{Dunlop_2021_primer}
{Dunlop}, J.~S., {Abraham}, R.~G., {Ashby}, M. L.~N., {et~al.} 2021, 1837

\bibitem[{Egami {et~al.}(2005)Egami, Kneib, Rieke, Ellis, Richard, Rigby,
  Papovich, Stark, Santos, Huang, Dole, Le~Floc'h, \&
  P{\'e}rez-Gonz{\'a}lez}]{egami_spitzer_2005}
Egami, E., Kneib, J.-P., Rieke, G.~H., {et~al.} 2005, The Astrophysical
  Journal, 618, L5, \dodoi{10.1086/427550}

\bibitem[{{Euclid Collaboration} {et~al.}(2022){Euclid Collaboration},
  {Moneti}, {McCracken}, {Shuntov}, {Kauffmann}, {Capak}, {Davidzon}, {Ilbert},
  {Scarlata}, {Toft}, {Weaver}, {Chary}, {Cuby}, {Faisst}, {Masters},
  {McPartland}, {Mobasher}, {Sanders}, {Scaramella}, {Stern}, {Szapudi},
  {Teplitz}, {Zalesky}, {Amara}, {Auricchio}, {Bodendorf}, {Bonino},
  {Branchini}, {Brau-Nogue}, {Brescia}, {Brinchmann}, {Capobianco}, {Carbone},
  {Carretero}, {Castander}, {Castellano}, {Cavuoti}, {Cimatti}, {Cledassou},
  {Congedo}, {Conselice}, {Conversi}, {Copin}, {Corcione}, {Costille},
  {Cropper}, {Da Silva}, {Degaudenzi}, {Douspis}, {Dubath}, {Duncan}, {Dupac},
  {Dusini}, {Farrens}, {Ferriol}, {Fosalba}, {Frailis}, {Franceschi}, {Fumana},
  {Garilli}, {Gillis}, {Giocoli}, {Granett}, {Grazian}, {Grupp}, {Haugan},
  {Hoekstra}, {Holmes}, {Hormuth}, {Hudelot}, {Jahnke}, {Kermiche},
  {Kiessling}, {Kilbinger}, {Kitching}, {Kohley}, {K{\"u}mmel}, {Kunz},
  {Kurki-Suonio}, {Ligori}, {Lilje}, {Lloro}, {Maiorano}, {Mansutti},
  {Marggraf}, {Markovic}, {Marulli}, {Massey}, {Maurogordato}, {Meneghetti},
  {Merlin}, {Meylan}, {Moresco}, {Moscardini}, {Munari}, {Niemi}, {Padilla},
  {Paltani}, {Pasian}, {Pedersen}, {Pires}, {Poncet}, {Popa}, {Pozzetti},
  {Raison}, {Rebolo}, {Rhodes}, {Rix}, {Roncarelli}, {Rossetti}, {Saglia},
  {Schneider}, {Secroun}, {Seidel}, {Serrano}, {Sirignano}, {Sirri}, {Stanco},
  {Tallada-Cresp{\'\i}}, {Taylor}, {TerOBOBeno}, {Toledo-Moreo},
  {Torradeflot}, {Wang}, {Welikala}, {Weller}, {Zamorani}, {Zoubian},
  {Andreon}, {Bardelli}, {Camera}, {Graci{\'a}-Carpio}, {Medinaceli}, {Mei},
  {Polenta}, {Romelli}, {Sureau}, {Tenti}, {Vassallo}, {Zacchei}, {Zucca},
  {Baccigalupi}, {Balaguera-Antol{\'\i}nez}, {Bernardeau}, {Biviano},
  {Bolzonella}, {Bozzo}, {Burigana}, {Cabanac}, {Cappi}, {Carvalho}, {Casas},
  {Castignani}, {Colodro-Conde}, {Coupon}, {Courtois}, {Di Ferdinando},
  {Farina}, {Finelli}, {Flose-Reimberg}, {Fotopoulou}, {Galeotta}, {Ganga},
  {Garcia-Bellido}, {Gaztanaga}, {Gozaliasl}, {Hook}, {Joachimi}, {Kansal},
  {Keihanen}, {Kirkpatrick}, {Lindholm}, {Mainetti}, {Maino}, {Maoli},
  {Martinelli}, {Martinet}, {Maturi}, {Metcalf}, {Morgante}, {Morisset},
  {Nucita}, {Patrizii}, {Potter}, {Renzi}, {Riccio}, {S{\'a}nchez}, {Sapone},
  {Schirmer}, {Schultheis}, {Scottez}, {Sefusatti}, {Teyssier}, {Tubio},
  {Tutusaus}, {Valiviita}, {Viel}, \& {Hildebrandt}}]{Euclid_2022_irac}
{Euclid Collaboration}, {Moneti}, A., {McCracken}, H.~J., {et~al.} 2022, \aap,
  658, A126, \dodoi{10.1051/0004-6361/202142361}

\bibitem[{Faisst {et~al.}(2016)Faisst, Capak, Hsieh, Laigle, Salvato, Tasca,
  Cassata, Davidzon, Ilbert, Le~F{\`e}vre, Masters, McCracken, Steinhardt,
  Silverman, de~Barros, Hasinger, \& Scoville}]{faisst_coherent_2016}
Faisst, A.~L., Capak, P., Hsieh, B.~C., {et~al.} 2016, The Astrophysical
  Journal, 821, 122, \dodoi{10.3847/0004-637X/821/2/122}

\bibitem[{Finkelstein {et~al.}(2012{\natexlab{a}})Finkelstein, Papovich,
  Salmon, Finlator, Dickinson, Ferguson, Giavalisco, Koekemoer, Reddy, Bassett,
  Conselice, Dunlop, Faber, Grogin, Hathi, Kocevski, Lai, Lee, McLure,
  Mobasher, \& Newman}]{finkelstein_candels_2012-1}
Finkelstein, S.~L., Papovich, C., Salmon, B., {et~al.} 2012{\natexlab{a}}, The
  Astrophysical Journal, 756, 164, \dodoi{10.1088/0004-637X/756/2/164}

\bibitem[{Finkelstein {et~al.}(2012{\natexlab{b}})Finkelstein, Papovich, Ryan,
  Pawlik, Dickinson, Ferguson, Finlator, Koekemoer, Giavalisco, Cooray, Dunlop,
  Faber, Grogin, Kocevski, \& Newman}]{finkelstein_candels_2012}
Finkelstein, S.~L., Papovich, C., Ryan, R.~E., {et~al.} 2012{\natexlab{b}}, The
  Astrophysical Journal, 758, 93, \dodoi{10.1088/0004-637X/758/2/93}

\bibitem[{Finkelstein {et~al.}(2013)Finkelstein, Papovich, Dickinson, Song,
  Tilvi, Koekemoer, Finkelstein, Mobasher, Ferguson, Giavalisco, Reddy, Ashby,
  Dekel, Fazio, Fontana, Grogin, Huang, Kocevski, Rafelski, Weiner, \&
  Willner}]{finkelstein_galaxy_2013}
Finkelstein, S.~L., Papovich, C., Dickinson, M., {et~al.} 2013, Nature, 502,
  524, \dodoi{10.1038/nature12657}

\bibitem[{{Finkelstein} {et~al.}(2023){Finkelstein}, {Bagley}, {Ferguson},
  {Wilkins}, {Kartaltepe}, {Papovich}, {Yung}, {Haro}, {Behroozi}, {Dickinson},
  {Kocevski}, {Koekemoer}, {Larson}, {Le Bail}, {Morales},
  {P{\'e}rez-Gonz{\'a}lez}, {Burgarella}, {Dav{\'e}}, {Hirschmann},
  {Somerville}, {Wuyts}, {Bromm}, {Casey}, {Fontana}, {Fujimoto}, {Gardner},
  {Giavalisco}, {Grazian}, {Grogin}, {Hathi}, {Hutchison}, {Jha}, {Jogee},
  {Kewley}, {Kirkpatrick}, {Long}, {Lotz}, {Pentericci}, {Pierel}, {Pirzkal},
  {Ravindranath}, {Ryan}, {Trump}, {Yang}, {Bhatawdekar}, {Bisigello}, {Buat},
  {Calabr{\`o}}, {Castellano}, {Cleri}, {Cooper}, {Croton}, {Daddi}, {Dekel},
  {Elbaz}, {Franco}, {Gawiser}, {Holwerda}, {Huertas-Company}, {Jaskot},
  {Leung}, {Lucas}, {Mobasher}, {Pandya}, {Tacchella}, {Weiner}, \&
  {Zavala}}]{Finkelstein_2023_CEERS}
{Finkelstein}, S.~L., {Bagley}, M.~B., {Ferguson}, H.~C., {et~al.} 2023, \apjl,
  946, L13, \dodoi{10.3847/2041-8213/acade4}

\bibitem[{{Fujimoto} {et~al.}(2023){Fujimoto}, {Arrabal Haro}, {Dickinson},
  {Finkelstein}, {Kartaltepe}, {Larson}, {Burgarella}, {Bagley}, {Behroozi},
  {Chworowsky}, {Hirschmann}, {Trump}, {Wilkins}, {Yung}, {Koekemoer},
  {Papovich}, {Pirzkal}, {Ferguson}, {Fontana}, {Grogin}, {Grazian}, {Kewley},
  {Kocevski}, {Lotz}, {Pentericci}, {Ravindranath}, {Somerville}, {Wilkins},
  {Amor{\'\i}n}, {Backhaus}, {Calabr{\`o}}, {Casey}, {Cooper}, {Fern{\'a}ndez},
  {Franco}, {Giavalisco}, {Hathi}, {Harish}, {Hutchison}, {Iyer}, {Jung},
  {Lucas}, \& {Zavala}}]{Fujimoto_2023_11z9_13GLXCEERS}
{Fujimoto}, S., {Arrabal Haro}, P., {Dickinson}, M., {et~al.} 2023, \apjl, 949,
  L25, \dodoi{10.3847/2041-8213/acd2d9}

\bibitem[{{Giavalisco} {et~al.}(2004){Giavalisco}, {Ferguson}, {Koekemoer},
  {Dickinson}, {Alexander}, {Bauer}, {Bergeron}, {Biagetti}, {Brandt},
  {Casertano}, {Cesarsky}, {Chatzichristou}, {Conselice}, {Cristiani}, {Da
  Costa}, {Dahlen}, {de Mello}, {Eisenhardt}, {Erben}, {Fall}, {Fassnacht},
  {Fosbury}, {Fruchter}, {Gardner}, {Grogin}, {Hook}, {Hornschemeier}, {Idzi},
  {Jogee}, {Kretchmer}, {Laidler}, {Lee}, {Livio}, {Lucas}, {Madau},
  {Mobasher}, {Moustakas}, {Nonino}, {Padovani}, {Papovich}, {Park},
  {Ravindranath}, {Renzini}, {Richardson}, {Riess}, {Rosati}, {Schirmer},
  {Schreier}, {Somerville}, {Spinrad}, {Stern}, {Stiavelli}, {Strolger},
  {Urry}, {Vandame}, {Williams}, \& {Wolf}}]{Giavalisco_2004_goods}
{Giavalisco}, M., {Ferguson}, H.~C., {Koekemoer}, A.~M., {et~al.} 2004, \apjl,
  600, L93, \dodoi{10.1086/379232}

\bibitem[{Gonz{\'a}lez {et~al.}(2014)Gonz{\'a}lez, Bouwens, Illingworth,
  Labb{\'e}, Oesch, Franx, \& Magee}]{gonzalez_slow_2014}
Gonz{\'a}lez, V., Bouwens, R., Illingworth, G., {et~al.} 2014, The
  Astrophysical Journal, 781, 34, \dodoi{10.1088/0004-637X/781/1/34}

\bibitem[{{Grogin} {et~al.}(2011){Grogin}, {Kocevski}, {Faber}, {Ferguson},
  {Koekemoer}, {Riess}, {Acquaviva}, {Alexander}, {Almaini}, {Ashby}, {Barden},
  {Bell}, {Bournaud}, {Brown}, {Caputi}, {Casertano}, {Cassata}, {Castellano},
  {Challis}, {Chary}, {Cheung}, {Cirasuolo}, {Conselice}, {Roshan Cooray},
  {Croton}, {Daddi}, {Dahlen}, {Dav{\'e}}, {de Mello}, {Dekel}, {Dickinson},
  {Dolch}, {Donley}, {Dunlop}, {Dutton}, {Elbaz}, {Fazio}, {Filippenko},
  {Finkelstein}, {Fontana}, {Gardner}, {Garnavich}, {Gawiser}, {Giavalisco},
  {Grazian}, {Guo}, {Hathi}, {H{\"a}ussler}, {Hopkins}, {Huang}, {Huang},
  {Jha}, {Kartaltepe}, {Kirshner}, {Koo}, {Lai}, {Lee}, {Li}, {Lotz}, {Lucas},
  {Madau}, {McCarthy}, {McGrath}, {McIntosh}, {McLure}, {Mobasher},
  {Moustakas}, {Mozena}, {Nandra}, {Newman}, {Niemi}, {Noeske}, {Papovich},
  {Pentericci}, {Pope}, {Primack}, {Rajan}, {Ravindranath}, {Reddy}, {Renzini},
  {Rix}, {Robaina}, {Rodney}, {Rosario}, {Rosati}, {Salimbeni}, {Scarlata},
  {Siana}, {Simard}, {Smidt}, {Somerville}, {Spinrad}, {Straughn}, {Strolger},
  {Telford}, {Teplitz}, {Trump}, {van der Wel}, {Villforth}, {Wechsler},
  {Weiner}, {Wiklind}, {Wild}, {Wilson}, {Wuyts}, {Yan}, \&
  {Yun}}]{Grogin_2011_candles}
{Grogin}, N.~A., {Kocevski}, D.~D., {Faber}, S.~M., {et~al.} 2011, \apjs, 197,
  35, \dodoi{10.1088/0067-0049/197/2/35}

\bibitem[{Guaita {et~al.}(2015)Guaita, Melinder, Hayes, {\"O}stlin, Gonzalez,
  Micheva, Adamo, Mas-Hesse, Sandberg, Ot{\'\i}-Floranes, Schaerer, Verhamme,
  Freeland, Orlitov{\'a}, Laursen, Cannon, Duval, Rivera-Thorsen, Herenz,
  Kunth, Atek, Puschnig, Gruyters, \& Pardy}]{guaita_lyman_2015}
Guaita, L., Melinder, J., Hayes, M., {et~al.} 2015, Astronomy and Astrophysics,
  576, A51, \dodoi{10.1051/0004-6361/201425053}

\bibitem[{{Hall}(2023)}]{Hall_2023_GP_JWST_Nat}
{Hall}, S. 2023, \nat, 613, 425, \dodoi{10.1038/d41586-023-00064-7}

\bibitem[{{Hayes} {et~al.}(2023){Hayes}, {Runnholm}, {Scarlata}, {Gronke}, \&
  {Rivera-Thorsen}}]{Hayes_2023}
{Hayes}, M.~J., {Runnholm}, A., {Scarlata}, C., {Gronke}, M., \&
  {Rivera-Thorsen}, T.~E. 2023, \mnras, 520, 5903,
  \dodoi{10.1093/mnras/stad477}

\bibitem[{{Henry} {et~al.}(2015){Henry}, {Scarlata}, {Martin}, \&
  {Erb}}]{Henry_2015_GP}
{Henry}, A., {Scarlata}, C., {Martin}, C.~L., \& {Erb}, D. 2015, \apj, 809, 19,
  \dodoi{10.1088/0004-637X/809/1/19}

\bibitem[{{Hsiao} {et~al.}(2023){Hsiao}, {Coe}, {Abdurro'uf}, {Whitler},
  {Jung}, {Khullar}, {Meena}, {Dayal}, {Barrow}, {Santos-Olmsted}, {Casselman},
  {Vanzella}, {Nonino}, {Jim{\'e}nez-Teja}, {Oguri}, {Stark}, {Furtak},
  {Zitrin}, {Adamo}, {Brammer}, {Bradley}, {Diego}, {Zackrisson},
  {Finkelstein}, {Windhorst}, {Bhatawdekar}, {Hutchison}, {Broadhurst},
  {Dimauro}, {Andrade-Santos}, {Eldridge}, {Acebron}, {Avila}, {Bayliss},
  {Ben{\'\i}tez}, {Binggeli}, {Bolan}, {Brada{\v{c}}}, {Carnall}, {Conselice},
  {Donahue}, {Frye}, {Fujimoto}, {Henry}, {James}, {Kassin}, {Kewley},
  {Larson}, {Lauer}, {Law}, {Mahler}, {Mainali}, {McCandliss}, {Nicholls},
  {Pirzkal}, {Postman}, {Rigby}, {Ryan}, {Senchyna}, {Sharon}, {Shimizu},
  {Strait}, {Tang}, {Trenti}, {Vikaeus}, \& {Welch}}]{Hsiao_2023_z10GLX_merger}
{Hsiao}, T. Y.-Y., {Coe}, D., {Abdurro'uf}, {et~al.} 2023, \apjl, 949, L34,
  \dodoi{10.3847/2041-8213/acc94b}

\bibitem[{Hu {et~al.}(2010)Hu, Cowie, Barger, Capak, Kakazu, \&
  Trouille}]{hu_atlas_2010}
Hu, E.~M., Cowie, L.~L., Barger, A.~J., {et~al.} 2010, The Astrophysical
  Journal, 725, 394, \dodoi{10.1088/0004-637X/725/1/394}

\bibitem[{{Inoue}(2011)}]{Inoue_2011_nebular}
{Inoue}, A.~K. 2011, \mnras, 415, 2920,
  \dodoi{10.1111/j.1365-2966.2011.18906.x}

\bibitem[{{Inoue} {et~al.}(2014){Inoue}, {Shimizu}, {Iwata}, \&
  {Tanaka}}]{Inoue_2014_igm}
{Inoue}, A.~K., {Shimizu}, I., {Iwata}, I., \& {Tanaka}, M. 2014, \mnras, 442,
  1805, \dodoi{10.1093/mnras/stu936}

\bibitem[{{Jarvis} {et~al.}(2013){Jarvis}, {Bonfield}, {Bruce}, {Geach},
  {McAlpine}, {McLure}, {Gonz{\'a}lez-Solares}, {Irwin}, {Lewis}, {Yoldas},
  {Andreon}, {Cross}, {Emerson}, {Dalton}, {Dunlop}, {Hodgkin}, {Le},
  {Karouzos}, {Meisenheimer}, {Oliver}, {Rawlings}, {Simpson}, {Smail},
  {Smith}, {Sullivan}, {Sutherland}, {White}, \& {Zwart}}]{Jarvis_2013_vista}
{Jarvis}, M.~J., {Bonfield}, D.~G., {Bruce}, V.~A., {et~al.} 2013, \mnras, 428,
  1281, \dodoi{10.1093/mnras/sts118}

\bibitem[{Jiang {et~al.}(2015)Jiang, Finlator, Cohen, Egami, Windhorst, Fan,
  Dav{\'e}, Kashikawa, Mechtley, Ouchi, Shimasaku, \&
  Cl{\'e}ment}]{jiang_physical_2015}
Jiang, L., Finlator, K., Cohen, S.~H., {et~al.} 2015, The Astrophysical
  Journal, 816, 16, \dodoi{10.3847/0004-637X/816/1/16}

\bibitem[{Jiang {et~al.}(2017)Jiang, Shen, Bian, Zheng, Wu, Oyarz{\'u}n, Blanc,
  Fan, Ho, Infante, Wang, Wu, Mateo, Bailey, Crane, Olszewski, Shectman,
  Thompson, \& Walker}]{jiang_magellan_2017}
Jiang, L., Shen, Y., Bian, F., {et~al.} 2017, The Astrophysical Journal, 846,
  134, \dodoi{10.3847/1538-4357/aa8561}

\bibitem[{Jiang {et~al.}(2018)Jiang, Wu, Bian, Chiang, Ho, Shen, Zheng, Bailey,
  Blanc, Crane, Fan, Mateo, Olszewski, Oyarz{\'u}n, Wang, \&
  Wu}]{jiang_giant_2018}
Jiang, L., Wu, J., Bian, F., {et~al.} 2018, Nature Astronomy, 2, 962,
  \dodoi{10.1038/s41550-018-0587-9}

\bibitem[{Jiang {et~al.}(2022)Jiang, Ning, Fan, Ho, Luo, Wang, Wu, Wu, Yang, \&
  Zheng}]{jiang_definitive_2022}
Jiang, L., Ning, Y., Fan, X., {et~al.} 2022, Nature Astronomy, 6, 850,
  \dodoi{10.1038/s41550-022-01708-w}

\bibitem[{Jung {et~al.}(2018)Jung, Finkelstein, Livermore, Dickinson, Larson,
  Papovich, Song, Tilvi, \& Wold}]{jung_texas_2018}
Jung, I., Finkelstein, S.~L., Livermore, R.~C., {et~al.} 2018, The
  Astrophysical Journal, 864, 103, \dodoi{10.3847/1538-4357/aad686}

\bibitem[{Jung {et~al.}(2022)Jung, Papovich, Finkelstein, Simons,
  Estrada-Carpenter, Backhaus, Cleri, Finlator, Giavalisco, Ji, Matharu,
  Momcheva, Straughn, \& Trump}]{jung_clear_2022}
Jung, I., Papovich, C., Finkelstein, S.~L., {et~al.} 2022, The Astrophysical
  Journal, 933, 87, \dodoi{10.3847/1538-4357/ac6fe7}

\bibitem[{Karman {et~al.}(2017)Karman, Caputi, Caminha, Gronke, Grillo,
  Balestra, Rosati, Vanzella, Coe, Dijkstra, Koekemoer, McLeod, Mercurio, \&
  Nonino}]{karman_muse_2017}
Karman, W., Caputi, K.~I., Caminha, G.~B., {et~al.} 2017, Astronomy \&
  Astrophysics, 599, A28, \dodoi{10.1051/0004-6361/201629055}

\bibitem[{Kashikawa {et~al.}(2006)Kashikawa, Shimasaku, Malkan, Doi, Matsuda,
  Ouchi, Taniguchi, Ly, Nagao, Iye, Motohara, Murayama, Murozono, Nariai, Ohta,
  Okamura, Sasaki, Shioya, \& Umemura}]{kashikawa_2006_subaru}
Kashikawa, N., Shimasaku, K., Malkan, M.~A., {et~al.} 2006, The Astrophysical
  Journal, 648, 7, \dodoi{10.1086/504966}

\bibitem[{Kawamata {et~al.}(2015)Kawamata, Ishigaki, Shimasaku, Oguri, \&
  Ouchi}]{kawamata_sizes_2015}
Kawamata, R., Ishigaki, M., Shimasaku, K., Oguri, M., \& Ouchi, M. 2015, The
  Astrophysical Journal, 804, 103, \dodoi{10.1088/0004-637X/804/2/103}

\bibitem[{Kobayashi {et~al.}(2016)Kobayashi, Murata, Koekemoer, Murayama,
  Taniguchi, Kajisawa, Shioya, Scoville, Nagao, \&
  Capak}]{kobayashi_morphological_2016}
Kobayashi, M. A.~R., Murata, K.~L., Koekemoer, A.~M., {et~al.} 2016, The
  Astrophysical Journal, 819, 25, \dodoi{10.3847/0004-637X/819/1/25}

\bibitem[{{Kocevski} {et~al.}(2018){Kocevski}, {Hasinger}, {Brightman},
  {Nandra}, {Georgakakis}, {Cappelluti}, {Civano}, {Li}, {Li}, {Aird},
  {Alexander}, {Almaini}, {Brusa}, {Buchner}, {Comastri}, {Conselice},
  {Dickinson}, {Finoguenov}, {Gilli}, {Koekemoer}, {Miyaji}, {Mullaney},
  {Papovich}, {Rosario}, {Salvato}, {Silverman}, {Somerville}, \&
  {Ueda}}]{Kocevski_2018_XUDS}
{Kocevski}, D.~D., {Hasinger}, G., {Brightman}, M., {et~al.} 2018, \apjs, 236,
  48, \dodoi{10.3847/1538-4365/aab9b4}

\bibitem[{{Koekemoer} {et~al.}(2011){Koekemoer}, {Faber}, {Ferguson}, {Grogin},
  {Kocevski}, {Koo}, {Lai}, {Lotz}, {Lucas}, {McGrath}, {Ogaz}, {Rajan},
  {Riess}, {Rodney}, {Strolger}, {Casertano}, {Castellano}, {Dahlen},
  {Dickinson}, {Dolch}, {Fontana}, {Giavalisco}, {Grazian}, {Guo}, {Hathi},
  {Huang}, {van der Wel}, {Yan}, {Acquaviva}, {Alexander}, {Almaini}, {Ashby},
  {Barden}, {Bell}, {Bournaud}, {Brown}, {Caputi}, {Cassata}, {Challis},
  {Chary}, {Cheung}, {Cirasuolo}, {Conselice}, {Roshan Cooray}, {Croton},
  {Daddi}, {Dav{\'e}}, {de Mello}, {de Ravel}, {Dekel}, {Donley}, {Dunlop},
  {Dutton}, {Elbaz}, {Fazio}, {Filippenko}, {Finkelstein}, {Frazer}, {Gardner},
  {Garnavich}, {Gawiser}, {Gruetzbauch}, {Hartley}, {H{\"a}ussler},
  {Herrington}, {Hopkins}, {Huang}, {Jha}, {Johnson}, {Kartaltepe},
  {Khostovan}, {Kirshner}, {Lani}, {Lee}, {Li}, {Madau}, {McCarthy},
  {McIntosh}, {McLure}, {McPartland}, {Mobasher}, {Moreira}, {Mortlock},
  {Moustakas}, {Mozena}, {Nandra}, {Newman}, {Nielsen}, {Niemi}, {Noeske},
  {Papovich}, {Pentericci}, {Pope}, {Primack}, {Ravindranath}, {Reddy},
  {Renzini}, {Rix}, {Robaina}, {Rosario}, {Rosati}, {Salimbeni}, {Scarlata},
  {Siana}, {Simard}, {Smidt}, {Snyder}, {Somerville}, {Spinrad}, {Straughn},
  {Telford}, {Teplitz}, {Trump}, {Vargas}, {Villforth}, {Wagner}, {Wandro},
  {Wechsler}, {Weiner}, {Wiklind}, {Wild}, {Wilson}, {Wuyts}, \&
  {Yun}}]{Koekemoer_2011_candles}
{Koekemoer}, A.~M., {Faber}, S.~M., {Ferguson}, H.~C., {et~al.} 2011, \apjs,
  197, 36, \dodoi{10.1088/0067-0049/197/2/36}

\bibitem[{Larson {et~al.}(2022)Larson, Finkelstein, Hutchison, Papovich,
  Bagley, Dickinson, Rojas-Ruiz, Ferguson, Jung, Giavalisco, Grazian,
  Pentericci, \& Tacchella}]{larson_searching_2022}
Larson, R.~L., Finkelstein, S.~L., Hutchison, T.~A., {et~al.} 2022, The
  Astrophysical Journal, 930, 104, \dodoi{10.3847/1538-4357/ac5dbd}

\bibitem[{{Lawrence} {et~al.}(2007){Lawrence}, {Warren}, {Almaini}, {Edge},
  {Hambly}, {Jameson}, {Lucas}, {Casali}, {Adamson}, {Dye}, {Emerson},
  {Foucaud}, {Hewett}, {Hirst}, {Hodgkin}, {Irwin}, {Lodieu}, {McMahon},
  {Simpson}, {Smail}, {Mortlock}, \& {Folger}}]{Warren_2007_UKIDSS}
{Lawrence}, A., {Warren}, S.~J., {Almaini}, O., {et~al.} 2007, \mnras, 379,
  1599, \dodoi{10.1111/j.1365-2966.2007.12040.x}

\bibitem[{{Leethochawalit} {et~al.}(2023){Leethochawalit}, {Trenti}, {Santini},
  {Yang}, {Merlin}, {Castellano}, {Fontana}, {Treu}, {Mason}, {Glazebrook},
  {Jones}, {Vulcani}, {Nanayakkara}, {Marchesini}, {Mascia}, {Morishita},
  {Roberts-Borsani}, {Bonchi}, {Paris}, {Boyett}, {Strait}, {Calabr{\`o}},
  {Pentericci}, {Bradac}, {Wang}, \&
  {Scarlata}}]{Leethochawalit_2023_jwst_dropout}
{Leethochawalit}, N., {Trenti}, M., {Santini}, P., {et~al.} 2023, \apjl, 942,
  L26, \dodoi{10.3847/2041-8213/ac959b}

\bibitem[{Liu {et~al.}(2017)Liu, Mutch, Poole, Angel, Duffy, Geil, Mesinger, \&
  Wyithe}]{liu_dark-ages_2017}
Liu, C., Mutch, S.~J., Poole, G.~B., {et~al.} 2017, Monthly Notices of the
  Royal Astronomical Society, 465, 3134, \dodoi{10.1093/mnras/stw2912}

\bibitem[{{Luo} {et~al.}(2017){Luo}, {Brandt}, {Xue}, {Lehmer}, {Alexander},
  {Bauer}, {Vito}, {Yang}, {Basu-Zych}, {Comastri}, {Gilli}, {Gu},
  {Hornschemeier}, {Koekemoer}, {Liu}, {Mainieri}, {Paolillo}, {Ranalli},
  {Rosati}, {Schneider}, {Shemmer}, {Smail}, {Sun}, {Tozzi}, {Vignali}, \&
  {Wang}}]{Luo_2017_cdfs}
{Luo}, B., {Brandt}, W.~N., {Xue}, Y.~Q., {et~al.} 2017, \apjs, 228, 2,
  \dodoi{10.3847/1538-4365/228/1/2}

\bibitem[{{Mascia} {et~al.}(2023){Mascia}, {Pentericci}, {Calabr{\`o}}, {Treu},
  {Santini}, {Yang}, {Napolitano}, {Roberts-Borsani}, {Bergamini}, {Grillo},
  {Rosati}, {Vulcani}, {Castellano}, {Boyett}, {Fontana}, {Glazebrook},
  {Henry}, {Mason}, {Merlin}, {Morishita}, {Nanayakkara}, {Paris}, {Roy},
  {Williams}, {Wang}, {Brammer}, {Brada{\v{c}}}, {Chen}, {Kelly}, {Koekemoer},
  {Trenti}, \& {Windhorst}}]{Mascia_2023_glass}
{Mascia}, S., {Pentericci}, L., {Calabr{\`o}}, A., {et~al.} 2023, \aap, 672,
  A155, \dodoi{10.1051/0004-6361/202345866}

\bibitem[{Mateo {et~al.}(2012)Mateo, Bailey, Crane, Shectman, Thompson,
  Roederer, Bigelow, \& Gunnels}]{mateo_m2fs_2012}
Mateo, M., Bailey, J.~I., Crane, J., {et~al.} 2012, 84464Y,
  \dodoi{10.1117/12.926448}

\bibitem[{{Matthee} {et~al.}(2023){Matthee}, {Mackenzie}, {Simcoe}, {Kashino},
  {Lilly}, {Bordoloi}, \& {Eilers}}]{Matthee_2023_EIGER}
{Matthee}, J., {Mackenzie}, R., {Simcoe}, R.~A., {et~al.} 2023, \apj, 950, 67,
  \dodoi{10.3847/1538-4357/acc846}

\bibitem[{McLeod {et~al.}(2016)McLeod, McLure, \& Dunlop}]{mcleod_z_2016}
McLeod, D.~J., McLure, R.~J., \& Dunlop, J.~S. 2016, Monthly Notices of the
  Royal Astronomical Society, 459, 3812, \dodoi{10.1093/mnras/stw904}

\bibitem[{{Naidu} {et~al.}(2022){Naidu}, {Oesch}, {van Dokkum}, {Nelson},
  {Suess}, {Brammer}, {Whitaker}, {Illingworth}, {Bouwens}, {Tacchella},
  {Matthee}, {Allen}, {Bezanson}, {Conroy}, {Labbe}, {Leja}, {Leonova},
  {Magee}, {Price}, {Setton}, {Strait}, {Stefanon}, {Toft}, {Weaver}, \&
  {Weibel}}]{Naidu_2022}
{Naidu}, R.~P., {Oesch}, P.~A., {van Dokkum}, P., {et~al.} 2022, \apjl, 940,
  L14, \dodoi{10.3847/2041-8213/ac9b22}

\bibitem[{{Ning} {et~al.}(2022){Ning}, {Cai}, {Jiang}, {Lin}, {Fu}, \&
  {Spinoso}}]{Ning_2022_JWST}
{Ning}, Y., {Cai}, Z., {Jiang}, L., {et~al.} 2022, arXiv e-prints,
  arXiv:2211.13620.
\newblock \doarXiv{2211.13620}

\bibitem[{Ning {et~al.}(2022)Ning, Jiang, Zheng, \& Wu}]{ning_magellan_2022}
Ning, Y., Jiang, L., Zheng, Z.-Y., \& Wu, J. 2022, The Astrophysical Journal,
  926, 230, \dodoi{10.3847/1538-4357/ac4268}

\bibitem[{Ning {et~al.}(2020)Ning, Jiang, Zheng, Wu, Bian, Egami, Fan, Ho,
  Shen, Wang, \& Wu}]{ning_magellan_2020}
Ning, Y., Jiang, L., Zheng, Z.-Y., {et~al.} 2020, The Astrophysical Journal,
  903, 4, \dodoi{10.3847/1538-4357/abb705}

\bibitem[{{Noll} {et~al.}(2009){Noll}, {Burgarella}, {Giovannoli}, {Buat},
  {Marcillac}, \& {Mu{\~n}oz-Mateos}}]{cigale:2009}
{Noll}, S., {Burgarella}, D., {Giovannoli}, E., {et~al.} 2009, \aap, 507, 1793,
  \dodoi{10.1051/0004-6361/200912497}

\bibitem[{{Oesch} {et~al.}(2023){Oesch}, {Brammer}, {Naidu}, {Bouwens},
  {Chisholm}, {Illingworth}, {Matthee}, {Nelson}, {Qin}, {Reddy}, {Shapley},
  {Shivaei}, {van Dokkum}, {Weibel}, {Whitaker}, {Wuyts}, {Covelo-Paz},
  {Endsley}, {Fudamoto}, {Giovinazzo}, {Herard-Demanche}, {Kerutt},
  {Kramarenko}, {Labbe}, {Leonova}, {Lin}, {Magee}, {Marchesini}, {Maseda},
  {Mason}, {Matharu}, {Meyer}, {Neufeld}, {Prieto Lyon}, {Schaerer}, {Sharma},
  {Shuntov}, {Smit}, {Stefanon}, {Wyithe}, \& {Xiao}}]{Oesch_2023_grism_survey}
{Oesch}, P.~A., {Brammer}, G., {Naidu}, R.~P., {et~al.} 2023, \mnras, 525,
  2864, \dodoi{10.1093/mnras/stad2411}

\bibitem[{{Ono} {et~al.}(2010){Ono}, {Ouchi}, {Shimasaku}, {Akiyama}, {Dunlop},
  {Farrah}, {Lee}, {McLure}, {Okamura}, \& {Yoshida}}]{Ono_2010}
{Ono}, Y., {Ouchi}, M., {Shimasaku}, K., {et~al.} 2010, \mnras, 402, 1580,
  \dodoi{10.1111/j.1365-2966.2009.16034.x}

\bibitem[{Ono {et~al.}(2012)Ono, Ouchi, Mobasher, Dickinson, Penner, Shimasaku,
  Weiner, Kartaltepe, Nakajima, Nayyeri, Stern, Kashikawa, \&
  Spinrad}]{ono_spectroscopic_2012}
Ono, Y., Ouchi, M., Mobasher, B., {et~al.} 2012, The Astrophysical Journal,
  744, 83, \dodoi{10.1088/0004-637X/744/2/83}

\bibitem[{Ono {et~al.}(2018)Ono, Ouchi, Harikane, Toshikawa, Rauch, Yuma,
  Sawicki, Shibuya, Shimasaku, Oguri, Willott, Akhlaghi, Akiyama, Coupon,
  Kashikawa, Komiyama, Konno, Lin, Matsuoka, Miyazaki, Nagao, Nakajima,
  Silverman, Tanaka, Taniguchi, \& Wang}]{ono_great_2018_GOLDRUSH}
Ono, Y., Ouchi, M., Harikane, Y., {et~al.} 2018, Publications of the
  Astronomical Society of Japan, 70, \dodoi{10.1093/pasj/psx103}

\bibitem[{Onoue {et~al.}(2017)Onoue, Kashikawa, Willott, Hibon, Im, Furusawa,
  Harikane, Imanishi, Ishikawa, Kikuta, Matsuoka, Nagao, Niino, Ono, Ouchi,
  Tanaka, Tang, Toshikawa, \& Uchiyama}]{onoue_minor_2017}
Onoue, M., Kashikawa, N., Willott, C.~J., {et~al.} 2017, The Astrophysical
  Journal, 847, L15, \dodoi{10.3847/2041-8213/aa8cc6}

\bibitem[{Oyarz{\'u}n {et~al.}(2017)Oyarz{\'u}n, Blanc, Gonz{\'a}lez, Mateo, \&
  Bailey}]{oyarzun_comprehensive_2017}
Oyarz{\'u}n, G.~A., Blanc, G.~A., Gonz{\'a}lez, V., Mateo, M., \& Bailey, J.~I.
  2017, The Astrophysical Journal, 843, 133, \dodoi{10.3847/1538-4357/aa7552}

\bibitem[{Oyarz{\'u}n {et~al.}(2016)Oyarz{\'u}n, Blanc, Gonz{\'a}lez, Mateo,
  Iii, Finkelstein, Lira, Crane, \& Olszewski}]{oyarzun_how_2016}
Oyarz{\'u}n, G.~A., Blanc, G.~A., Gonz{\'a}lez, V., {et~al.} 2016, The
  Astrophysical Journal, 821, L14, \dodoi{10.3847/2041-8205/821/1/L14}

\bibitem[{{Peng} {et~al.}(2002){Peng}, {Ho}, {Impey}, \& {Rix}}]{galfit:2002}
{Peng}, C.~Y., {Ho}, L.~C., {Impey}, C.~D., \& {Rix}, H.-W. 2002, \aj, 124,
  266, \dodoi{10.1086/340952}

\bibitem[{{Peng} {et~al.}(2010){Peng}, {Ho}, {Impey}, \& {Rix}}]{galfit:2010}
---. 2010, \aj, 139, 2097, \dodoi{10.1088/0004-6256/139/6/2097}

\bibitem[{Pentericci {et~al.}(2011)Pentericci, Fontana, Vanzella, Castellano,
  Grazian, Dijkstra, Boutsia, Cristiani, Dickinson, Giallongo, Giavalisco,
  Maiolino, Moorwood, Paris, \& Santini}]{pentericci_spectroscopic_2011}
Pentericci, L., Fontana, A., Vanzella, E., {et~al.} 2011, The Astrophysical
  Journal, 743, 132, \dodoi{10.1088/0004-637X/743/2/132}

\bibitem[{{Prieto-Lyon} {et~al.}(2023){Prieto-Lyon}, {Mason}, {Mascia},
  {Merlin}, {Roy}, {Henry}, {Roberts-Borsani}, {Morishita}, {Wang}, {Boyett},
  {Bolan}, {Bradac}, {Castellano}, {Mercurio}, {Nanayakkara}, {Paris},
  {Pentericci}, {Scarlata}, {Trenti}, {Treu}, \&
  {Vanzella}}]{Prieto-Lyon_2023_NIRSpec}
{Prieto-Lyon}, G., {Mason}, C., {Mascia}, S., {et~al.} 2023, \apj, 956, 136,
  \dodoi{10.3847/1538-4357/acf715}

\bibitem[{{Rhoads} {et~al.}(2023){Rhoads}, {Wold}, {Harish}, {Kim}, {Pharo},
  {Malhotra}, {Gabrielpillai}, {Jiang}, \& {Yang}}]{Rhoads_2023_GP_JWST}
{Rhoads}, J.~E., {Wold}, I. G.~B., {Harish}, S., {et~al.} 2023, \apjl, 942,
  L14, \dodoi{10.3847/2041-8213/acaaaf}

\bibitem[{{Roberts-Borsani} {et~al.}(2023){Roberts-Borsani}, {Treu}, {Chen},
  {Morishita}, {Vanzella}, {Zitrin}, {Bergamini}, {Castellano}, {Fontana},
  {Glazebrook}, {Grillo}, {Kelly}, {Merlin}, {Nanayakkara}, {Paris}, {Rosati},
  {Yang}, {Acebron}, {Bonchi}, {Boyett}, {Brada{\v{c}}}, {Brammer},
  {Broadhurst}, {Calabr{\'o}}, {Diego}, {Dressler}, {Furtak}, {Filippenko},
  {Henry}, {Koekemoer}, {Leethochawalit}, {Malkan}, {Mason}, {Mercurio},
  {Metha}, {Pentericci}, {Pierel}, {Rieck}, {Roy}, {Santini}, {Strait},
  {Strausbaugh}, {Trenti}, {Vulcani}, {Wang}, {Wang}, \&
  {Windhorst}}]{Roberts_Borsani_2023_z10GLX}
{Roberts-Borsani}, G., {Treu}, T., {Chen}, W., {et~al.} 2023, \nat, 618, 480,
  \dodoi{10.1038/s41586-023-05994-w}

\bibitem[{Roberts-Borsani {et~al.}(2016)Roberts-Borsani, Bouwens, Oesch, Labbe,
  Smit, Illingworth, Dokkum, Holden, Gonzalez, Stefanon, Holwerda, \&
  Wilkins}]{roberts-borsani_z_2016}
Roberts-Borsani, G.~W., Bouwens, R.~J., Oesch, P.~A., {et~al.} 2016, The
  Astrophysical Journal, 823, 143, \dodoi{10.3847/0004-637X/823/2/143}

\bibitem[{Robertson {et~al.}(2015)Robertson, Ellis, Furlanetto, \&
  Dunlop}]{robertson_cosmic_2015}
Robertson, B.~E., Ellis, R.~S., Furlanetto, S.~R., \& Dunlop, J.~S. 2015, The
  Astrophysical Journal, 802, L19, \dodoi{10.1088/2041-8205/802/2/L19}

\bibitem[{Rodriguez-Gomez {et~al.}(2015)Rodriguez-Gomez, Genel, Vogelsberger,
  Sijacki, Pillepich, Sales, Torrey, Snyder, Nelson, Springel, Ma, \&
  Hernquist}]{Rodriguez_2015}
Rodriguez-Gomez, V., Genel, S., Vogelsberger, M., {et~al.} 2015, Monthly
  Notices of the Royal Astronomical Society, 449, 49,
  \dodoi{10.1093/mnras/stv264}

\bibitem[{{Roy} {et~al.}(2023){Roy}, {Henry}, {Treu}, {Jones}, {Prieto-Lyon},
  {Mason}, {Heckman}, {Nanayakkara}, {Pentericci}, {Mascia}, {Brada{\v{c}}},
  {Vanzella}, {Scarlata}, {Boyett}, {Trenti}, \& {Wang}}]{Roy_2023}
{Roy}, N., {Henry}, A., {Treu}, T., {et~al.} 2023, \apjl, 952, L14,
  \dodoi{10.3847/2041-8213/acdbce}

\bibitem[{{Schenker} {et~al.}(2014){Schenker}, {Ellis}, {Konidaris}, \&
  {Stark}}]{Schenker_2014}
{Schenker}, M.~A., {Ellis}, R.~S., {Konidaris}, N.~P., \& {Stark}, D.~P. 2014,
  \apj, 795, 20, \dodoi{10.1088/0004-637X/795/1/20}

\bibitem[{Schenker {et~al.}(2012)Schenker, Stark, Ellis, Robertson, Dunlop,
  McLure, Kneib, \& Richard}]{schenker_keck_2012}
Schenker, M.~A., Stark, D.~P., Ellis, R.~S., {et~al.} 2012, The Astrophysical
  Journal, 744, 179, \dodoi{10.1088/0004-637X/744/2/179}

\bibitem[{Schmidt {et~al.}(2016)Schmidt, Treu, Brada{\v c}, Vulcani, Huang,
  Hoag, Maseda, Guaita, Pentericci, Brammer, Dijkstra, Dressler, Fontana,
  Henry, Jones, Mason, Trenti, \& Wang}]{schmidt_grism_2016}
Schmidt, K.~B., Treu, T., Brada{\v c}, M., {et~al.} 2016, The Astrophysical
  Journal, 818, 38, \dodoi{10.3847/0004-637X/818/1/38}

\bibitem[{{Shen} {et~al.}(2020){Shen}, {Hopkins}, {Faucher-Gigu{\`e}re},
  {Alexander}, {Richards}, {Ross}, \& {Hickox}}]{Shen_2020_sed}
{Shen}, X., {Hopkins}, P.~F., {Faucher-Gigu{\`e}re}, C.-A., {et~al.} 2020,
  \mnras, 495, 3252, \dodoi{10.1093/mnras/staa1381}

\bibitem[{Shibuya {et~al.}(2015)Shibuya, Ouchi, \&
  Harikane}]{shibuya_morphologies_2015}
Shibuya, T., Ouchi, M., \& Harikane, Y. 2015, The Astrophysical Journal
  Supplement Series, 219, 15, \dodoi{10.1088/0067-0049/219/2/15}

\bibitem[{Shibuya {et~al.}(2016)Shibuya, Ouchi, Kubo, \&
  Harikane}]{shibuya_morphologies_2016}
Shibuya, T., Ouchi, M., Kubo, M., \& Harikane, Y. 2016, The Astrophysical
  Journal, 821, 72, \dodoi{10.3847/0004-637X/821/2/72}

\bibitem[{{Shimasaku} {et~al.}(2006){Shimasaku}, {Kashikawa}, {Doi}, {Ly},
  {Malkan}, {Matsuda}, {Ouchi}, {Hayashino}, {Iye}, {Motohara}, {Murayama},
  {Nagao}, {Ohta}, {Okamura}, {Sasaki}, {Shioya}, \&
  {Taniguchi}}]{Shimasaku_2006_skewness}
{Shimasaku}, K., {Kashikawa}, N., {Doi}, M., {et~al.} 2006, \pasj, 58, 313,
  \dodoi{10.1093/pasj/58.2.313}

\bibitem[{Song {et~al.}(2016)Song, Finkelstein, Livermore, Capak, Dickinson, \&
  Fontana}]{song_keckmosfire_2016}
Song, M., Finkelstein, S.~L., Livermore, R.~C., {et~al.} 2016, The
  Astrophysical Journal, 826, 113, \dodoi{10.3847/0004-637X/826/2/113}

\bibitem[{{Stark} {et~al.}(1992){Stark}, {Gammie}, {Wilson}, {Bally}, {Linke},
  {Heiles}, \& {Hurwitz}}]{Stark_1992_NH}
{Stark}, A.~A., {Gammie}, C.~F., {Wilson}, R.~W., {et~al.} 1992, \apjs, 79, 77,
  \dodoi{10.1086/191645}

\bibitem[{Stark {et~al.}(2010)Stark, Ellis, Chiu, Ouchi, \&
  Bunker}]{stark_keck_2010}
Stark, D.~P., Ellis, R.~S., Chiu, K., Ouchi, M., \& Bunker, A. 2010, Monthly
  Notices of the Royal Astronomical Society, 408, 1628,
  \dodoi{10.1111/j.1365-2966.2010.17227.x}

\bibitem[{{Stark} {et~al.}(2011){Stark}, {Ellis}, \& {Ouchi}}]{Stark_2011_Keck}
{Stark}, D.~P., {Ellis}, R.~S., \& {Ouchi}, M. 2011, \apjl, 728, L2,
  \dodoi{10.1088/2041-8205/728/1/L2}

\bibitem[{Stark {et~al.}(2011)Stark, Ellis, \& Ouchi}]{stark_keck_2011}
Stark, D.~P., Ellis, R.~S., \& Ouchi, M. 2011, The Astrophysical Journal, 728,
  L2, \dodoi{10.1088/2041-8205/728/1/L2}

\bibitem[{Stark {et~al.}(2013)Stark, Schenker, Ellis, Robertson, McLure, \&
  Dunlop}]{stark_keck_2013}
Stark, D.~P., Schenker, M.~A., Ellis, R., {et~al.} 2013, The Astrophysical
  Journal, 763, 129, \dodoi{10.1088/0004-637X/763/2/129}

\bibitem[{Sun {et~al.}(2023)Sun, Ho, Zhuang, Ma, Chen, \&
  Li}]{sun_2023_structure}
Sun, W., Ho, L.~C., Zhuang, M.-Y., {et~al.} 2023, The Structure and Morphology
  of Galaxies during the Epoch of Reionization Revealed by JWST.
\newblock \doarXiv{2308.09076}

\bibitem[{{Tacchella} {et~al.}(2023){Tacchella}, {Johnson}, {Robertson},
  {Carniani}, {D'Eugenio}, {Kumari}, {Maiolino}, {Nelson}, {Suess},
  {{\"U}bler}, {Williams}, {Adebusola}, {Alberts}, {Arribas}, {Bhatawdekar},
  {Bonaventura}, {Bowler}, {Bunker}, {Cameron}, {Curti}, {Egami}, {Eisenstein},
  {Frye}, {Hainline}, {Helton}, {Ji}, {Looser}, {Lyu}, {Perna}, {Rawle},
  {Rieke}, {Rieke}, {Saxena}, {Sandles}, {Shivaei}, {Simmonds}, {Sun},
  {Willmer}, {Willott}, \& {Witstok}}]{Tacchella_2023_NIRSpec}
{Tacchella}, S., {Johnson}, B.~D., {Robertson}, B.~E., {et~al.} 2023, \mnras,
  522, 6236, \dodoi{10.1093/mnras/stad1408}

\bibitem[{{Tang} {et~al.}(2023){Tang}, {Stark}, {Chen}, {Mason}, {Topping},
  {Endsley}, {Senchyna}, {Plat}, {Lu}, {Whitler}, {Robertson}, \&
  {Charlot}}]{Tang_2023_ceersspec}
{Tang}, M., {Stark}, D.~P., {Chen}, Z., {et~al.} 2023, \mnras, 526, 1657,
  \dodoi{10.1093/mnras/stad2763}

\bibitem[{Tilvi {et~al.}(2014)Tilvi, Papovich, Finkelstein, Long, Song,
  Dickinson, Ferguson, Koekemoer, Giavalisco, \& Mobasher}]{tilvi_rapid_2014}
Tilvi, V., Papovich, C., Finkelstein, S.~L., {et~al.} 2014, The Astrophysical
  Journal, 794, 5, \dodoi{10.1088/0004-637X/794/1/5}

\bibitem[{Toshikawa {et~al.}(2012)Toshikawa, Kashikawa, Ota, Morokuma, Shibuya,
  Hayashi, Nagao, Jiang, Malkan, Egami, Shimasaku, Motohara, \&
  Ishizaki}]{toshikawa_discovery_2012}
Toshikawa, J., Kashikawa, N., Ota, K., {et~al.} 2012, The Astrophysical
  Journal, 750, 137, \dodoi{10.1088/0004-637X/750/2/137}

\bibitem[{{Vanden Berk} {et~al.}(2001){Vanden Berk}, {Richards}, {Bauer},
  {Strauss}, {Schneider}, {Heckman}, {York}, {Hall}, {Fan}, {Knapp},
  {Anderson}, {Annis}, {Bahcall}, {Bernardi}, {Briggs}, {Brinkmann}, {Brunner},
  {Burles}, {Carey}, {Castander}, {Connolly}, {Crocker}, {Csabai}, {Doi},
  {Finkbeiner}, {Friedman}, {Frieman}, {Fukugita}, {Gunn}, {Hennessy},
  {Ivezi{\'c}}, {Kent}, {Kunszt}, {Lamb}, {Leger}, {Long}, {Loveday}, {Lupton},
  {Meiksin}, {Merelli}, {Munn}, {Newberg}, {Newcomb}, {Nichol}, {Owen}, {Pier},
  {Pope}, {Rockosi}, {Schlegel}, {Siegmund}, {Smee}, {Snir}, {Stoughton},
  {Stubbs}, {SubbaRao}, {Szalay}, {Szokoly}, {Tremonti}, {Uomoto}, {Waddell},
  {Yanny}, \& {Zheng}}]{Berk_2001_sdss_composite_spec}
{Vanden Berk}, D.~E., {Richards}, G.~T., {Bauer}, A., {et~al.} 2001, \aj, 122,
  549, \dodoi{10.1086/321167}

\bibitem[{{Vanzella} {et~al.}(2009){Vanzella}, {Giavalisco}, {Dickinson},
  {Cristiani}, {Nonino}, {Kuntschner}, {Popesso}, {Rosati}, {Renzini}, {Stern},
  {Cesarsky}, {Ferguson}, \& {Fosbury}}]{Vanzella_2009_VLT}
{Vanzella}, E., {Giavalisco}, M., {Dickinson}, M., {et~al.} 2009, \apj, 695,
  1163, \dodoi{10.1088/0004-637X/695/2/1163}

\bibitem[{Vanzella {et~al.}(2014)Vanzella, Fontana, Pentericci, Castellano,
  Grazian, Giavalisco, Nonino, Cristiani, Zamorani, \&
  Vignali}]{vanzella_52_2014}
Vanzella, E., Fontana, A., Pentericci, L., {et~al.} 2014, Astronomy \&
  Astrophysics, 569, A78, \dodoi{10.1051/0004-6361/201424285}

\bibitem[{{Wang} {et~al.}(2023){Wang}, {Yang}, {Hennawi}, {Fan}, {Sun},
  {Champagne}, {Costa}, {Habouzit}, {Endsley}, {Li}, {Lin}, {Meyer},
  {Schindler}, {Wu}, {Ba{\~n}ados}, {Barth}, {Bhowmick}, {Bieri}, {Blecha},
  {Bosman}, {Cai}, {Colina}, {Connor}, {Davies}, {Decarli}, {De Rosa}, {Drake},
  {Egami}, {Eilers}, {Evans}, {Farina}, {Haiman}, {Jiang}, {Jin}, {Jun},
  {Kakiichi}, {Khusanova}, {Kulkarni}, {Li}, {Liu}, {Loiacono}, {Lupi},
  {Mazzucchelli}, {Onoue}, {Pudoka}, {Rojas-Ruiz}, {Shen}, {Strauss}, {Tee},
  {Trakhtenbrot}, {Trebitsch}, {Venemans}, {Volonteri}, {Walter}, {Xie}, {Yue},
  {Zhang}, {Zhang}, \& {Zou}}]{Wang_2023_ASPIRE}
{Wang}, F., {Yang}, J., {Hennawi}, J.~F., {et~al.} 2023, \apjl, 951, L4,
  \dodoi{10.3847/2041-8213/accd6f}

\bibitem[{Watson {et~al.}(2015)Watson, Christensen, Knudsen, Richard, Gallazzi,
  \& Micha{\l}owski}]{watson_dusty_2015}
Watson, D., Christensen, L., Knudsen, K.~K., {et~al.} 2015, Nature, 519, 327,
  \dodoi{10.1038/nature14164}

\bibitem[{Wilkins {et~al.}(2010)Wilkins, Bunker, Ellis, Stark, Stanway, Chiu,
  Lorenzoni, \& Jarvis}]{Wilkins_2010_hstLBGz7}
Wilkins, S.~M., Bunker, A.~J., Ellis, R.~S., {et~al.} 2010, Monthly Notices of
  the Royal Astronomical Society, 403, 938,
  \dodoi{10.1111/j.1365-2966.2009.16175.x}

\bibitem[{Wold {et~al.}(2022)Wold, Malhotra, Rhoads, Wang, Hu, Perez, Zheng,
  Khostovan, Walker, Barrientos, González-López, Harish, Infante, Jiang,
  Pharo, Moya-Sierralta, Bauer, Galaz, Valdes, \& Yang}]{Wold_2022_lager}
Wold, I. G.~B., Malhotra, S., Rhoads, J., {et~al.} 2022, The Astrophysical
  Journal, 927, 36, \dodoi{10.3847/1538-4357/ac4997}

\bibitem[{Wu {et~al.}(2020)Wu, Jiang, \& Ning}]{wu_diffuse_2020}
Wu, J., Jiang, L., \& Ning, Y. 2020, The Astrophysical Journal, 891, 105,
  \dodoi{10.3847/1538-4357/ab7333}

\bibitem[{{Yan} {et~al.}(2023){Yan}, {Sun}, {Ma}, \&
  {Ling}}]{Yan_2023_jwst_dropout_2}
{Yan}, H., {Sun}, B., {Ma}, Z., \& {Ling}, C. 2023, arXiv e-prints,
  arXiv:2311.15121, \dodoi{10.48550/arXiv.2311.15121}

\bibitem[{{Yang} {et~al.}(2017){Yang}, {Malhotra}, {Gronke}, {Rhoads},
  {Leitherer}, {Wofford}, {Jiang}, {Dijkstra}, {Tilvi}, \&
  {Wang}}]{Yang_2017_GP}
{Yang}, H., {Malhotra}, S., {Gronke}, M., {et~al.} 2017, \apj, 844, 171,
  \dodoi{10.3847/1538-4357/aa7d4d}

\bibitem[{Zheng {et~al.}(2017)Zheng, Wang, Rhoads, Infante, Malhotra, Hu,
  Walker, Jiang, Jiang, Hibon, Gonzalez, Kong, Zheng, Galaz, \&
  Barrientos}]{zheng_first_2017}
Zheng, Z.-Y., Wang, J., Rhoads, J., {et~al.} 2017, The Astrophysical Journal,
  842, L22, \dodoi{10.3847/2041-8213/aa794f}

\end{thebibliography}

\end{document}